\documentclass[10pt,twocolumn]{IEEEtran}

\usepackage{graphicx,subfigure}
\usepackage{psfig,epsfig,algorithm,amsmath,amssymb,color,subfigure,url}
\usepackage{caption}
\usepackage{multirow,bigstrut}
\begin{document}
\setlength{\arraycolsep}{0.03cm}
\newcommand{\xhat}{\hat{x}}
\newcommand{\xpred}{\hat{x}_{t|t-1}}
\newcommand{\xupd}{\hat{x}_{t|t}}
\newcommand{\Ppred}{P_{t|t-1}}
\newcommand{\ty}{\tilde{y}_t}
\newcommand{\tty}{\tilde{y}_{t,\text{res}}}
\newcommand{\tw}{\tilde{w}_t}
\newcommand{\ttw}{\tilde{w}_{t,f}}
\newcommand{\betahat}{\hat{\beta}}

\newcommand{\ypast}{y_{1:t-1}}
\newcommand{\sone}{S_{*}}
\newcommand{\sinf}{{S_{**}}}
\newcommand{\smax}{S_{\max}}
\newcommand{\smin}{S_{\min}}
\newcommand{\samax}{S_{a,\max}}
\newcommand{\Nhat}{{\hat{N}}}

\newcommand{\Dnum}{D_{num}}
\newcommand{\pss}{p^{**,i}}
\newcommand{\fr}{f_{r}^i}

\newcommand{\A}{{\cal A}}
\newcommand{\Z}{{\cal Z}}
\newcommand{\B}{{\cal B}}
\newcommand{\R}{{\cal R}}
\newcommand{\reg}{{\cal G}}
\newcommand{\const}{\mbox{const}}

\newcommand{\trace}{\mbox{tr}}

\newcommand{\hsim}{{\hspace{0.0cm} \sim  \hspace{0.0cm}}}
\newcommand{\he}{{\hspace{0.0cm} =  \hspace{0.0cm}}}

\newcommand{\vect}[2]{\left[\begin{array}{cccccc}
     #1 \\
     #2
   \end{array}
  \right]
  }

\newcommand{\matr}[2]{ \left[\begin{array}{cc}
     #1 \\
     #2
   \end{array}
  \right]
  }
\newcommand{\vc}[2]{\left[\begin{array}{c}
     #1 \\
     #2
   \end{array}
  \right]
  }

\newcommand{\gdot}{\dot{g}}
\newcommand{\Cdot}{\dot{C}}
\newcommand{\re}{\mathbb{R}}
\newcommand{\n}{{\cal N}}  
\newcommand{\N}{{\overrightarrow{\bf N}}}  
\newcommand{\chat}{\tilde{C}_t}
\newcommand{\chati}{\chat^i}

\newcommand{\cmin}{C^*_{min}}
\newcommand{\twi}{\tilde{w}_t^{(i)}}
\newcommand{\twj}{\tilde{w}_t^{(j)}}
\newcommand{\wi}{{w}_t^{(i)}}
\newcommand{\twio}{\tilde{w}_{t-1}^{(i)}}

\newcommand{\tWi}{\tilde{W}_n^{(m)}}
\newcommand{\tWj}{\tilde{W}_n^{(k)}}
\newcommand{\Wi}{{W}_n^{(m)}}
\newcommand{\tWio}{\tilde{W}_{n-1}^{(m)}}

\newcommand{\ds}{\displaystyle}

\newcommand{\SAR}{S$\!$A$\!$R }
\newcommand{\MAR}{MAR}
\newcommand{\MMRF}{MMRF}
\newcommand{\AR}{A$\!$R }
\newcommand{\GMRF}{G$\!$M$\!$R$\!$F }
\newcommand{\DTM}{D$\!$T$\!$M }
\newcommand{\MSE}{M$\!$S$\!$E }
\newcommand{\RCS}{R$\!$C$\!$S }
\newcommand{\uomega}{\underline{\omega}}
\newcommand{\y}{v}
\newcommand{\x}{w}
\newcommand{\lu}{\mu}
\newcommand{\g}{g}
\newcommand{\s}{{\bf s}}
\newcommand{\bft}{{\bf t}}
\newcommand{\refmap}{{\cal R}}
\newcommand{\totrefl}{{\cal E}}
\newcommand{\beq}{\begin{equation}}
\newcommand{\eeq}{\end{equation}}
\newcommand{\bdm}{\begin{displaymath}}
\newcommand{\edm}{\end{displaymath}}
\newcommand{\hatz}{\hat{z}}
\newcommand{\hatu}{\hat{u}}
\newcommand{\tilz}{\tilde{z}}
\newcommand{\tilu}{\tilde{u}}
\newcommand{\hhatz}{\hat{\hat{z}}}
\newcommand{\hhatu}{\hat{\hat{u}}}
\newcommand{\tilc}{\tilde{C}}
\newcommand{\hatc}{\hat{C}}
\newcommand{\tim}{n}

\newcommand{\ssp}{\renewcommand{\baselinestretch}{1.0}}
\newcommand{\defd}{\mbox{$\stackrel{\mbox{$\triangle$}}{=}$}}
\newcommand{\goes}{\rightarrow}
\newcommand{\tends}{\rightarrow}
\newcommand{\defn}{\triangleq} 
\newcommand{\se}{&=&}
\newcommand{\sdefn}{& \defn  &}
\newcommand{\sle}{& \le &}
\newcommand{\sge}{& \ge &}
\newcommand{\plusminus}{\stackrel{+}{-}}
\newcommand{\Ey}{E_{Y_{1:t}}}
\newcommand{\ey}{E_{Y_{1:t}}}

\newcommand{\equivto}{\mbox{~~~which is equivalent to~~~}}
\newcommand{\nonzero}{i:\pi^n(x^{(i)})>0}
\newcommand{\nonzeroc}{i:c(x^{(i)})>0}

\newcommand{\supn}{\sup_{\phi:||\phi||_\infty \le 1}}
\newtheorem{definition}{Definition}
\newtheorem{remark}{Remark}
\newtheorem{example}{Example}
\newtheorem{ass}{Assumption}
\newtheorem{proposition}{Proposition}

\newtheorem{fact}{Fact}
\newtheorem{heuristic}{Heuristic}
\newcommand{\eps}{\epsilon}
\newcommand{\bd}{\begin{definition}}
\newcommand{\ed}{\end{definition}}
\newcommand{\udq}{\underline{D_Q}}
\newcommand{\td}{\tilde{D}}
\newcommand{\epsinv}{\epsilon_{inv}}
\newcommand{\al}{\mathcal{A}}

\newcommand{\bfx} {\bf X}
\newcommand{\bfy} {\bf Y}
\newcommand{\bfz} {\bf Z}
\newcommand{\ddas}{\mbox{${d_1}^2({\bf X})$}}
\newcommand{\ddbs}{\mbox{${d_2}^2({\bfx})$}}
\newcommand{\dda}{\mbox{$d_1(\bfx)$}}
\newcommand{\ddb}{\mbox{$d_2(\bfx)$}}
\newcommand{\xinc}{{\bfx} \in \mbox{$C_1$}}
\newcommand{\eqa}{\stackrel{(a)}{=}}
\newcommand{\eqb}{\stackrel{(b)}{=}}
\newcommand{\eqe}{\stackrel{(e)}{=}}
\newcommand{\leqc}{\stackrel{(c)}{\le}}
\newcommand{\leqd}{\stackrel{(d)}{\le}}

\newcommand{\leqa}{\stackrel{(a)}{\le}}
\newcommand{\leqb}{\stackrel{(b)}{\le}}
\newcommand{\leqe}{\stackrel{(e)}{\le}}
\newcommand{\leqf}{\stackrel{(f)}{\le}}
\newcommand{\leqg}{\stackrel{(g)}{\le}}
\newcommand{\leqh}{\stackrel{(h)}{\le}}
\newcommand{\leqi}{\stackrel{(i)}{\le}}
\newcommand{\leqj}{\stackrel{(j)}{\le}}

\newcommand{\w}{{W^{LDA}}}
\newcommand{\halpha}{\hat{\alpha}}
\newcommand{\hsigma}{\hat{\sigma}}
\newcommand{\slmax}{\sqrt{\lambda_{max}}}
\newcommand{\slmin}{\sqrt{\lambda_{min}}}
\newcommand{\lmax}{\lambda_{max}}
\newcommand{\lmin}{\lambda_{min}}

\newcommand{\da} {\frac{\alpha}{\sigma}}
\newcommand{\chka} {\frac{\check{\alpha}}{\check{\sigma}}}
\newcommand{\sumo}{\sum _{\underline{\omega} \in \Omega}}
\newcommand{\distance}{d\{(\hatz _x, \hatz _y),(\tilz _x, \tilz _y)\}}
\newcommand{\col}{{\rm col}}
\newcommand{\rcs}{\sigma_0}
\newcommand{\CalR}{{\cal R}}
\newcommand{\df}{{\delta p}}
\newcommand{\dq}{{\delta q}}
\newcommand{\dZ}{{\delta Z}}
\newcommand{\pprime}{{\prime\prime}}

\newcommand{\vn}{N}

\newcommand{\bv}{\begin{vugraph}}
\newcommand{\ev}{\end{vugraph}}
\newcommand{\bi}{\begin{itemize}}
\newcommand{\ei}{\end{itemize}}
\newcommand{\ben}{\begin{enumerate}}
\newcommand{\een}{\end{enumerate}}
\newcommand{\be}{\protect\[}
\newcommand{\ee}{\protect\]}
\newcommand{\bean}{\begin{eqnarray*} }
\newcommand{\eean}{\end{eqnarray*} }
\newcommand{\bea}{\begin{eqnarray} }
\newcommand{\eea}{\end{eqnarray} }
\newcommand{\nn}{\nonumber}
\newcommand{\ba}{\begin{array} }
\newcommand{\ea}{\end{array} }
\newcommand{\ep}{\mbox{\boldmath $\epsilon$}}
\newcommand{\epp}{\mbox{\boldmath $\epsilon '$}}
\newcommand{\Lep}{\mbox{\LARGE $\epsilon_2$}}
\newcommand{\und}{\underline}
\newcommand{\pdif}[2]{\frac{\partial #1}{\partial #2}}
\newcommand{\odif}[2]{\frac{d #1}{d #2}}
\newcommand{\dt}[1]{\pdif{#1}{t}}
\newcommand{\urho}{\underline{\rho}}

\newcommand{\spc}{{\cal S}}
\newcommand{\tspc}{{\cal TS}}

\newcommand{\uv}{\underline{v}}
\newcommand{\us}{\underline{s}}
\newcommand{\uc}{\underline{c}}
\newcommand{\utheta}{\underline{\theta}^*}
\newcommand{\ualpha}{\underline{\alpha^*}}

\newcommand{\uxy}{\underline{x}^*}
\newcommand{\uxyj}{[x^{*}_j,y^{*}_j]}
\newcommand{\arcl}[1]{arclen(#1)}
\newcommand{\one}{{\mathbf{1}}}

\newcommand{\uxyjt}{\uxy_{j,t}}
\newcommand{\E}{\mathbb{E}}

\newcommand{\rhomat}{\left[\begin{array}{c}
                        \rho_3 \ \rho_4 \\
                        \rho_5 \ \rho_6
                        \end{array}
                   \right]}
\newcommand{\deltat}{\tau} 
\newcommand{\deltatt}{\Delta t_1}
\newcommand{\ceil}[1]{\ulcorner #1 \urcorner}

\newcommand{\xxi}{x^{(i)}}
\newcommand{\txi}{\tilde{x}^{(i)}}
\newcommand{\txj}{\tilde{x}^{(j)}}

\newcommand{\mi}[1]{{#1}^{(m,i)}}

\title{ReProCS: A Missing Link between Recursive Robust PCA and Recursive Sparse Recovery in Large but Correlated Noise}
\author{Chenlu Qiu and Namrata Vaswani
\thanks{The authors are with the Electrical and Computer Engineering Department, Iowa State University, Ames, IA 50010 USA (e-mail: chenlu@iastate.edu; namrata@iastate.edu). A part of this work was presented at Allerton 2010 and ISIT 2011 \cite{rrpcp_allerton,rrpcp_isit}. This research was partially supported by NSF grants ECCS-0725849 and CCF-0917015.
}
}

\maketitle

\vspace{-0.2in}

\begin{abstract}
This work studies the recursive robust principal components' analysis (PCA) problem. Here, ``robust" refers to robustness to both independent and correlated sparse outliers, although we focus on the latter. A key application where this problem occurs is in video surveillance where the goal is to separate a slowly changing background from moving foreground objects on-the-fly. The background sequence is well modeled as lying in a low dimensional subspace, that can gradually change over time, while the moving foreground objects constitute the correlated sparse outliers. In this and many other applications, the foreground is an outlier for PCA but is actually the ``signal of interest" for the application; where as the background is the corruption or noise. Thus our problem can also be interpreted as one of recursively recovering a time sequence of sparse signals in the presence of large but spatially correlated noise.

This work has two key contributions. First, we provide a new way of looking at this problem and show how a key part of our solution strategy involves solving a noisy compressive sensing (CS) problem. Second, we  show how we can utilize the correlation of the outliers to our advantage in order to even deal with very large support sized outliers. The main idea is as follows. The correlation model applied to the previous support estimate helps predict the current support. This prediction serves as ``partial support knowledge" for solving the modified-CS problem instead of CS. The support estimate of the modified-CS reconstruction is, in turn, used to update the correlation model parameters using a Kalman filter (or any adaptive filter). We call the resulting approach ``support-predicted modified-CS".
\end{abstract}

\vspace{-0.1in}

\newcommand{\Phat}{\hat{P}}
\section{Introduction} \label{intro}

Most high dimensional data often approximately lies in a lower dimensional subspace. Principal Components' Analysis (PCA) is a widely used dimension reduction technique that finds a small number of orthogonal basis vectors (principal components), along which most of the variability of the dataset lies. To be precise, for a given dimension, $r$, PCA finds the $r$-dimensional subspace that minimizes the mean squared error between data vectors and their $r$-dimensional projections \cite{PCA}. The subspace spanned by the principal components (PCs) is called the principal components' space (PC space). Often, for time series data, the PC space changes gradually over time. Updating the PC space recursively as more data comes in, without re-solving the entire PCA problem, is referred to as recursive PCA \cite{sequentialSVD}. 

Notice that to find an $r$ dimensional PC space, one needs at least $r$ data vectors, usually more. Thus, even for recursive PCA, the initial step needs to be a batch one or the initial PC space needs to be pre-specified.

It is well known that PCA is very sensitive to outliers. Computing the PCs in the presence of outliers is called robust PCA \cite{Roweis98emalgorithms,Torre03aframework,rpca,rpca_Chandrasekaran}.
Doing recursive PCA in the presence of outliers is referred to as recursive robust PCA \cite{sequentialSVD,ipca_weightedand,Li03anintegrated}.
``Outlier" is a loosely defined term that usually refers to any corruption that is not small compared to the true data vector and that occurs only occasionally. As suggested in \cite{error_correction_PCP_l1,rpca}, an outlier can be very nicely modeled as a sparse vector, i.e. a vector whose most elements are zero, while the few that are nonzero can have any magnitude. We will use this definition in this paper. In  \cite{error_correction_PCP_l1,rpca}, the outlier is modeled as being spatially and temporally independent. In most real applications, the time at which the outliers begin to occur is often random and independent of all past times. But once outliers begin to occur, for some time after that they are both spatially and temporally correlated. In this work, we focus on this case, i.e. on {\em recursive robust PCA that is robust to correlated sparse outliers}.

A key application where the robust PCA problem occurs is in video surveillance where the goal is to separate a slowly changing background from moving foreground objects \cite{Torre03aframework,rpca}. If we stack each video frame as a column vector, then the background is well modeled as lying in a low dimensional subspace that may gradually change over time, while the moving foreground objects constitute the sparse outliers \cite{error_correction_PCP_l1}\cite{rpca} which change in a correlated fashion over time. \emph{We will use this as the motivating application in this work.} Other important applications include sensor networks based detection and tracking of abnormal events such as forest fires or oil spills; or online detection of brain activation patterns from fMRI sequences (the ``active" part of the brain can be interpreted as a correlated sparse outlier). Clearly, in all these cases, one would need a real-time and fast solution and hence a recursive robust PCA solution is desirable.

The moving objects or the brain active regions or the oil spill region may be ``outliers" for the PCA problem, but in most cases, these are actually the ``signals of interest" whereas the background image is the noise. Also, all the above ``signals of interest"  are sparse vectors that change in a correlated fashion over time. Thus, this problem can also be re-interpreted as one of {\em recursively recovering a time sequence of correlated sparse signals, $S_t$, in the presence of large but ``low rank" noise, $L_t$}. 

\begin{definition}
The term ``low rank" vector means that the $n \times \tau$ matrix $L:=[L_{t-\tau+1},\dots L_t]$ has low-rank, i.e. its rank is less than $\min(n,\tau)$ for $\tau$ large enough. This would happen if $L_t$ is correlated enough to have a low rank covariance matrix and this matrix changes slowly over time. We make this precise below in Sec. \ref{intro_prob}.
\end{definition}

\subsection{Problem Definition}\label{intro_prob}
Our problem can be defined as follows.
The measurement vector at time, $t$, $M_t$, is an $n \times 1$ vector that satisfies
\begin{equation}
M_t = L_t  + S_t \label{eq1}
\end{equation}
where $S_t$ is a sparse vector, with support set denoted by $T_t$, and $L_t$ is a dense (non sparse) but ``low rank" vector. 
The support set of $S_t$, $T_t$, can be correlated over time and space. 

Suppose, we have a good estimate of the initial PC matrix, $\hat{P}_0 \approx P_0$. For $t>0$, our goal is to recursively keep estimating the sparse part, $S_t$, at each time, and to keep updating $\hat{P}_t$ every-so-often, by collecting the recent estimates of $L_t=M_t-S_t$.

To make things precise, we assume that $L_t$ satisfies
\begin{equation*}
L_t = U x_t
\end{equation*}
where $U$ is an {\em unknown} orthonormal matrix and $x_t$ is an $n \times 1$ sparse vector whose support changes every-so-often and whose elements are spatially uncorrelated. Let $N_t$ denote the support set of $x_t$. We assume that $N_t$ is piecewise constant with time. Thus, the columns of the sub-matrix, $P_t := (U)_{N_t}$, span the low dimensional subspace in which the current set of $L_t$'s lie and $L_t = P_t (x_t)_{N_t}$. We refer to $P_t$ as the principal components' (PC) matrix. Clearly, this is also piecewise constant with time.

Every $d$ time units, there are $k$ additions to the set $N_t$, or, equivalently, $k$ directions get added to $P_t$. When a new direction gets added, the magnitude of $x_t$ along it is initially small but gradually increases to a larger stable value. Also, the values of $x_t$ along $k$ existing directions gradually decays down to zero, i.e. the corresponding directions get slowly removed from $P_t$. We provide a generative model that satisfies these assumptions in the Appendix. It models $x_t$ (and hence $L_t$) as being piecewise stationary with short nonstationary transients between the pieces that occur whenever the support of $x_t$ changes.

In the video problem, the observed image is an {\em overlay} of the foreground and the background image. Denote the image, background image and foreground image written as a 1D vector by $M_t$, $L_t$ and $O_t$ respectively. Let $T_t$ denote the support of $O_t$ and let $T_t^c$ denote the complement set of $T_t$. By overlay, we mean that $(M_t)_{T_t} = (O_t)_{T_t}$ and $(M_t)_{T_t^c} = (L_t)_{T_t^c}$. We can rewrite this in the form of (\ref{eq1}) by defining $S_t$ as
\begin{equation}
(S_t)_i = \left\{ \begin{array}{ll}
(O_t-L_t)_i \  & \ \mbox{ $i \in T_t$}  \label{S_def}\\
 0           \ &  \ \mbox{ $i \in T_t^c$}
\end{array}
\right.
\end{equation}



\begin{figure*}
\centerline{
\subfigure[The nonzero elements of $S_t$ has large magnitude $100$]{
\psfig{file = 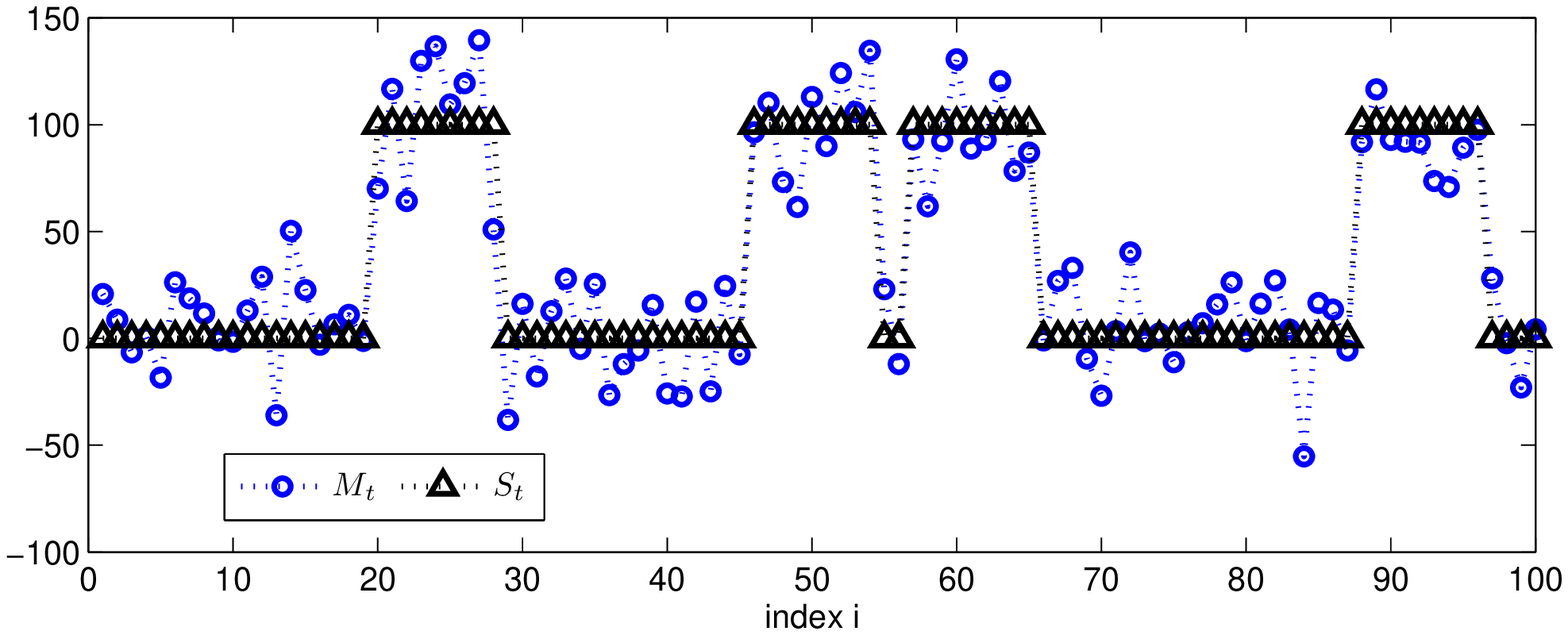, height=3cm,width=8cm} \label{largeS}
}
\hspace{-0.15in}
\subfigure[Recovery of Fig. \ref{largeS}]{
\psfig{file = 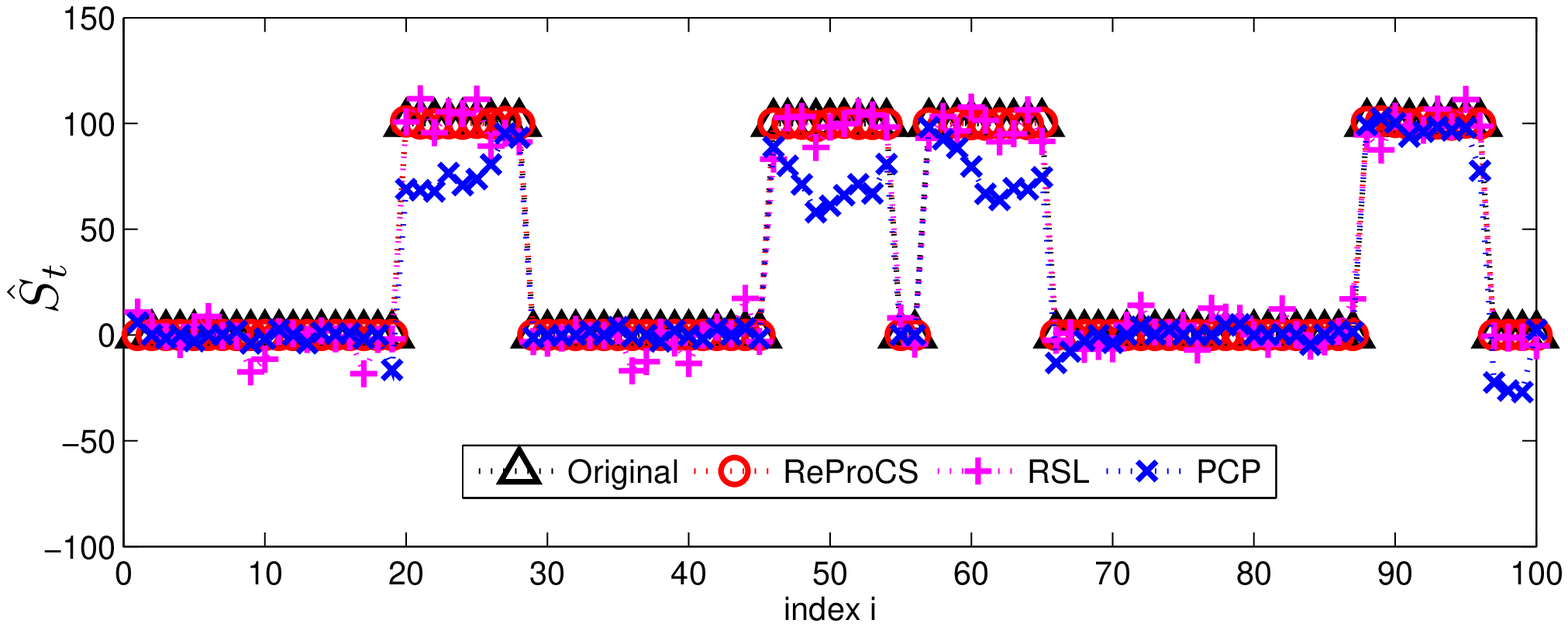, height=3cm,width=8cm} \label{largeShat}
}
}
\centerline{
\subfigure[The nonzero elements of $S_t$ has small magnitude $5$]{
\psfig{file = 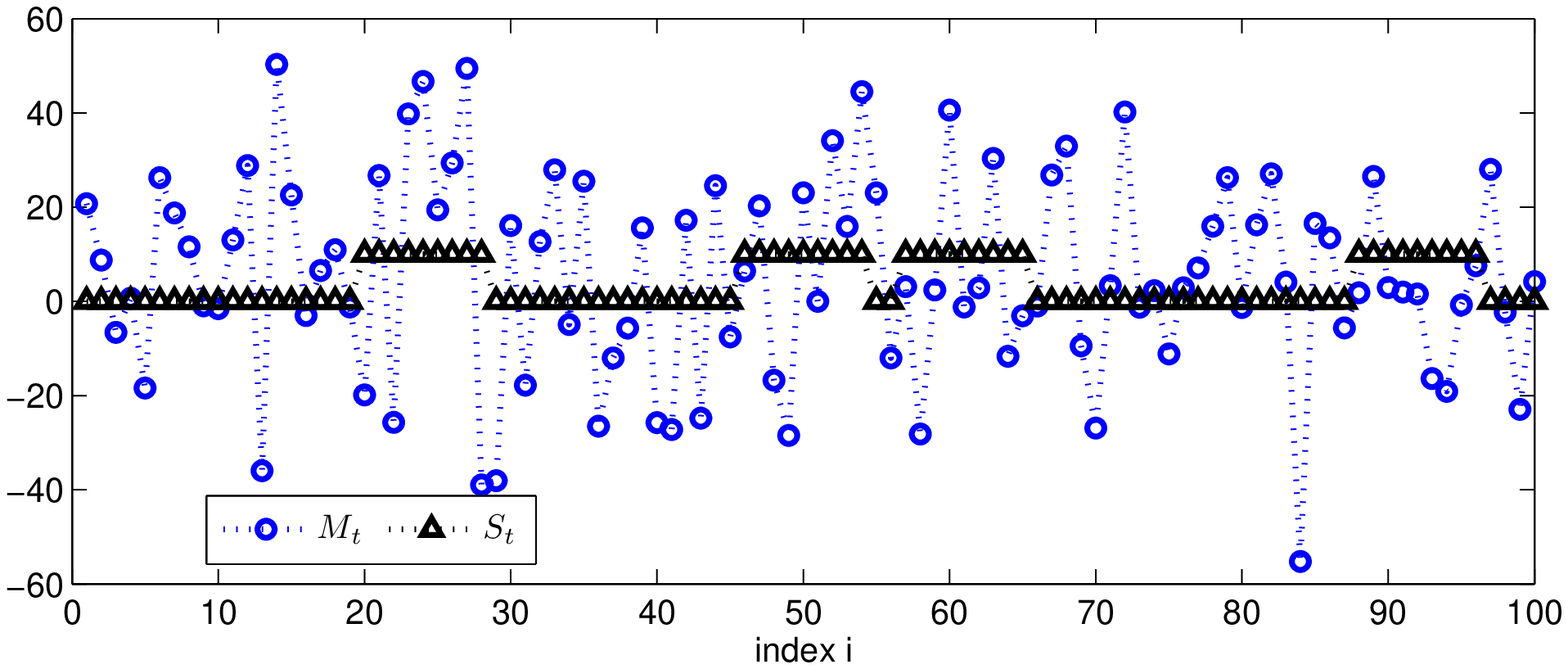, height=3cm,width=8cm} \label{smallS}
}
\hspace{-0.15in}
\subfigure[Recovery of Fig. \ref{smallS}]{
\psfig{file = 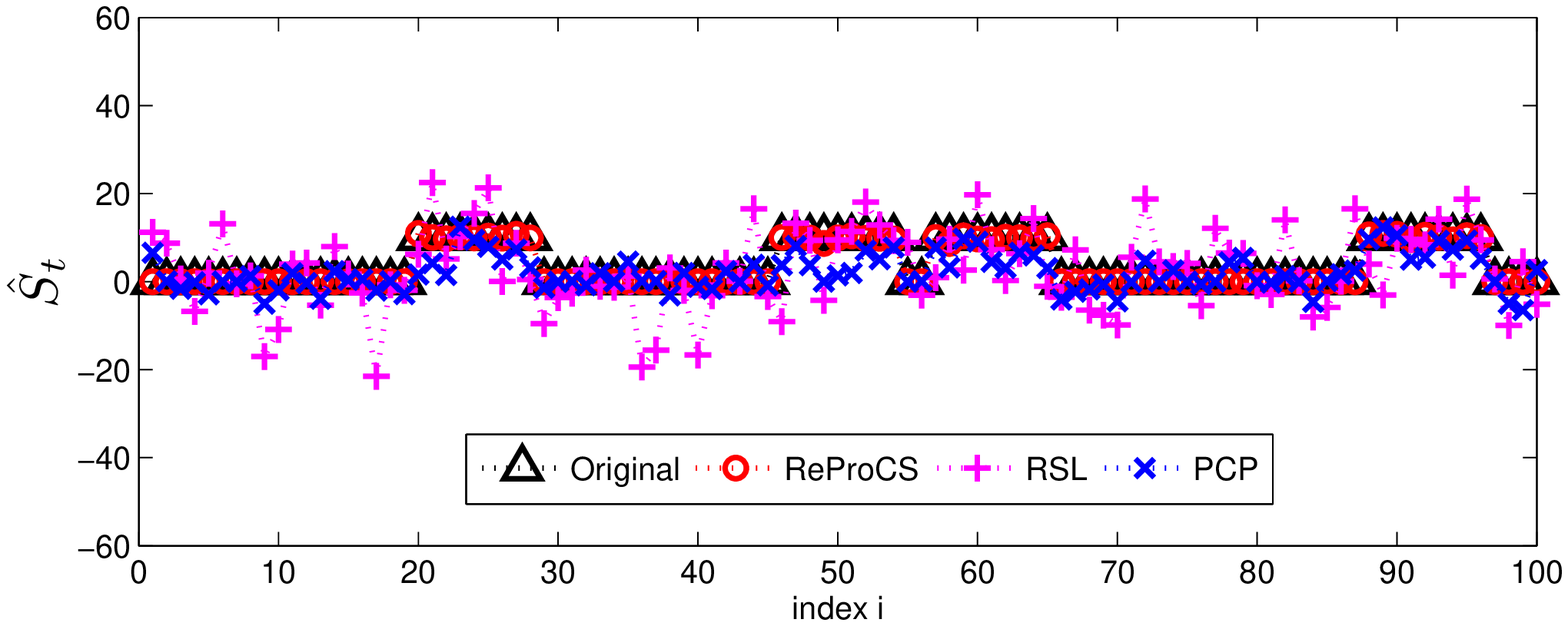, height=3cm,width=8cm} \label{smallShat}
}
}
\vspace{-0.15in}
\caption{\small{We plot $M_t$, $S_t$, and $\hat{S}_t$ at $t=t_0+5$ for the experiment described in Sec. \ref{expt_n100}. The sparse part $S_t$ has large magnitude in the top row and and small magnitude in the bottom row. As a result, RSL \cite{Torre03aframework} works in the first case but not the second one. The support sets of $S_t$'s are large and generated in a correlated fashion and so the sparse matrix $S=[S_1, \dots S_{t_0+200}]$ also has low rank (rank = 45\%). This is the reason PCP \cite{rpca} does not work in both cases (also see Table \ref{tabel_intro}, PCP does work in case of small magnitude but small support $S_t$'s).\label{expt1_intro}
\vspace{-0.15in}
}}
\end{figure*}

If we refer to the above problem as {\em recursive robust PCA}, then the low dimensional vector, $L_t$, is the ``signal of interest" while the correlated sparse vector, $S_t$, is the corruption (outlier). On the other hand, if we refer to it as {\em recursive sparse recovery in large but spatially correlated (low rank) noise, $L_t$}, then the correlated sparse vector, $S_t$, is the signal of interest, while $L_t$ is the corruption (large but low rank noise). An example of noise $L_t$ being much larger than signal $S_t$ is shown in Fig. \ref{smallS}. In the rest of the paper, we just refer to $S_t$ as the {\em sparse part} and $L_t$ as the {\em low rank part}.%

In this work, we focus on the $M_t = S_t + L_t$ case, but our proposed ideas will also apply if
\begin{equation}
M_t = \Psi S_t + L_t \label{general_model}
\end{equation}
where $\Psi$ can be a fat matrix. This would be the standard sparse recovery problem from a reduced number of measurements but with the difference that the measurements are corrupted by very large noise. Also, this allows the signal to be sparse in some other basis other than the canonical basis.
With the exception of \cite{error_correction_PCP_l1}, which can handle another kind of very large but ``structured" noise (the noise or outlier needs to be sparse), almost all other existing sparse recovery as well as recursive sparse recovery approaches for time sequences only work with small noise \cite{candesnoise,convexsparse,bpdn,SubSpaceCS,cosamp,wotao_yin,modelcs,LS-CS-residual,modcsjp} (surely none of these will work if the noise is significantly larger than the sparse part).
\subsection{Contributions and Paper Organization}
Our first contribution is to show how our problem can be reformulated as a sparse recovery / compressive sensing (CS) plus recursive PCA problem. We call our solution {\em Recursive Projected Compressive Sensing (ReProCS)}. Its block diagram is shown in Fig. \ref{chart1}. The key idea is as follows. Assume that the current PC matrix $P_t$ has been accurately estimated, i.e. we are given $\Phat_t \approx P_t$. We project the measurement vector, $M_t$, into the space perpendicular to $\Phat_t$ to get an $n-r$ dimensional projected measurement vector $y_t$. Here, $r$ is the rank of $\Phat_t$. This projection nullifies most of the contribution of $L_t$. It is assumed that this projection does not nullify any nonzero component of the sparse vectors, $S_t$. Recovering $S_t$ from $y_t$ now becomes a traditional noisy CS \cite{bpdn,candes,donoho} problem. The recovered $S_t$ can be used to estimate $L_t$ which can then be used to update $\Phat_t$ every-so-often (recursive PCA).
%
%
{\em In this work we misuse terminology a little and use ``compressive sensing" or ``CS" to refer to the $\ell_1$ minimization problem.}

Our second key contribution is to show how we can utilize the correlation of the sparse part, $S_t$, to our advantage to successfully recover $S_t$ even when its support size, $|T_t|$, increases for a given rank $r$ of $P_t$ and as a result the number of projected ``measurements" available for the CS step, $n-r$, become too small for CS to work. 
The correlated sparse outliers (e.g. moving foreground objects in video) can be interpreted as sparse signal sequences whose {\em support change over time is either slow;} {\em or quite often is not slow, but is still correlated, e.g. the support can ``move" or ``expand" or change according to some model, over time}. By using a model on how the objects move or deform, or other models on how the support changes, it is possible to obtain a support prediction that serves as an accurate ``partial support knowledge". We then tap into our recent work on modified-CS (modCS) which solves the sparse recovery problem with much fewer measurements when reliable support knowledge is available \cite{modcs_ts}. The support estimate of the modCS reconstruction can then be used to update the correlation model parameters using a Kalman filter or any other adaptive filter. We call this {\em support predicted modified-CS (supp-pred-modCS)}. Its block diagram shown in Fig. \ref{chart2}.
%

This paper is organized as follows. We discuss related work next and define notation in Sec. \ref{sec_notions}. The ReProCS idea is described in Sec. \ref{rrpcp_sec}. The supp-pred-modCS idea and the overall ReProCS(modCS) approach is developed in Sec. \ref{supp_rrpcp}.  We also discuss its stability over time.  In Sec. \ref{sec_updatePt}, we explain the recursive PCA algorithm which is based on the idea of \cite{sequentialSVD}. Simulation experiments and partly-real experiments (real background images with simulated foreground sparse image) evaluating the performance of ReProCS and ReProCS(modCS) and comparing them with other state-of-art methods --  two recursive robust PCA methods, \cite{Li03anintegrated} and \cite{sequentialSVD}, two batch robust PCA methods, principal components' pursuit (PCP) \cite{rpca} and robust subspace learning (RSL) \cite{Torre03aframework}, and with simple thresholding based background subtraction (for the video case) are given in Sec. \ref{results}. Conclusions, limitations and future work are given in Sec. \ref{sec_conclusion}.  
{\em Sections \ref{reprocs_assumption}, \ref{sec_stability} and \ref{sec_updatePt}  can be skipped on a quick reading. We mark them with a **.} 

{\em For ease of review, both our related work discussion and our experiments' section is very detailed. Some of this material can be shortened/removed eventually.}


\subsection{Related Work} 
There has been a large amount of work on robust PCA, e.g. \cite{Roweis98emalgorithms,Torre03aframework}, and recursive robust PCA e.g. \cite{sequentialSVD,ipca_weightedand,Li03anintegrated}. In most of these works, either the locations of the missing/corruped data points are assumed known \cite{sequentialSVD} (not a practical assumption); or they first detect the corrupted data points and then replace their values using nearby values \cite{ipca_weightedand}; or weight each data point in proportion to its reliability (thus soft-detecting and down-weighting the likely outliers) \cite{Torre03aframework,Li03anintegrated}; or just remove the entire outlier vector. Approaches like \cite{sequentialSVD} can be adapted to the case where the missing/corrupted data points are unknown by using the outlier detection approach from other works, e.g. from \cite{Li03anintegrated} (we refer to the resulting method as adapted-\cite{sequentialSVD}).
Detecting or soft-detecting outliers (sparse part $S_t$) as in \cite{ipca_weightedand,Torre03aframework,Li03anintegrated} is easy to do when their magnitude is large compared to that of $L_t$, but not when it is smaller, e.g. see Fig \ref{expt1_intro} and Table \ref{tabel_intro}. When the signal of interest is $S_t$ (the case of recursive sparse recovery in large but low rank noise), the most difficult situation is when nonzero elements of $S_t$ have small magnitude compared to those of $L_t$. Moreover, as we explain in Sec. \ref{results}, approaches such as \cite{Li03anintegrated} or adapted-\cite{sequentialSVD} also cannot work in case of too many outliers (large support size of $S_t$), e.g. see Table \ref{tabel_intro}. But ReProCS is able to successfully recover both small magnitude and fairly large support-sized $S_t$'s because it operates by first approximately nullifying $L_t$ and then recovering $S_t$ by solving a noisy CS problem, that enforces sparsity of the $S_t$'s.

In a series of recent works \cite{rpca,rpca_Chandrasekaran,error_correction_PCP}, an elegant solution has been proposed, that does not require a two step outlier location detection/correction process and also does not throw out the entire vector. It redefines batch robust PCA as a problem of separating a low rank matrix, $L := [L_1,\dots,L_t]$, from a sparse matrix, $S := [S_1,\dots,S_t]$, using the data matrix, $M := [M_1,\dots,M_t] = L + S$. It was shown in \cite{rpca} that one can recover $L$ and $S$ exactly by solving
\begin{equation}
\underset{L,S} {\min}\|L\|_* + \lambda\|S\|_1 \ \text{subject to}  \ \ L +S = M \label{PCP}
\end{equation}
where $\|L\|_*$ is the sum of singular values of $L$ while $\|S\|_1$ is the $\ell_1$ norm of $S$ seen as a long vector, provided that
\begin{itemize}
\item the singular vectors of $L$ are spread out enough (not sparse),
\item the support and signs of $S$ are uniformly random (thus it is not low rank) and
\item the rank of $L$ is sufficiently small for a given sparsity of $S$.
\end{itemize}
This was called {\em Principal Components' Pursuit (PCP)}. While PCP is an elegant idea, it has three important limitations.
\begin{itemize}
\item 
   Most importantly, PCP relies on the fact that the matrix $S:=[S_{1},\dots S_t]$ is sparse but full rank. But when the $S_t$'s are correlated and have a large support size, $S$ will also often be low rank. This is particularly easy to see for the case where the support sets, $T_t$, have large overlaps over time. For example, for the case of Fig \ref{expt1_intro}, the rank of $S$ is only $45$ (while $n=100$). A low rank $S$ makes it impossible for PCP to separate $S$ from $L$ and hence neither is recovered correctly. Also see Fig. \ref{expt1}.

\item A related issue is that PCP requires the rank of $L$ to be quite small for a given support size of $S_t$, $|T_t|$, e.g. see \cite[Table 1]{rpca}. But, real videos can have a lot of background variations, e.g. see Sec. \ref{expt_wave}, causing the rank of $L$ to be as much as 20\% of the image size. Also, the foreground can have large sized and multiple moving objects, making $|T_t|$ also quite large.

\item In many applications, e.g. in surveillance, one would like to obtain the estimates on-the-fly and quickly as a new frame comes in, rather than in a batch fashion.

\end{itemize}

Our proposed algorithms, ReProCS and ReProCS(modCS), address these drawbacks. Unlike PCP, (a) ReProCS does not need the $S_t$'s to be uncorrelated, and (b) it can recover $S_t$'s with larger support sizes [see Figs. \ref{expt1_intro}, \ref{expt1} and Table \ref{table1}]. Moreover, our second solution, ReProCS(modCS), utilizes the correlation of the $S_t$'s to its advantage in order to successfully recover $S_t$'s with significantly larger support sizes than what ReProCS can [see Figs \ref{expt3_two}, \ref{expt4_large}]. Finally, both ReProCS and  ReProCS(modCS) are recursive methods and hence they provide real-times estimates. Of course, our work and in fact any recursive robust PCA method, e.g. \cite{Li03anintegrated,sequentialSVD}, does need an initial estimate of the PC matrix which PCP or other batch methods, e.g. \cite{Torre03aframework}, do not need. In practice, as we explain later in Sec. \ref{sec_updatePt}, this is usually easy to obtain.


The static version of our problem is also somewhat related to that of \cite{decodinglp}, \cite{rpca_regression}, \cite{rpca_regression_sparse} in that all of these also try to cancel the ``low rank" part by projecting the original data vector into the perpendicular space of the tall matrix that spans the ``low rank" part. But the big difference is that in all these, this matrix is {\em known}. In our problem $P_t$ is unknown and can change with time, and we also show how we can utilize the correlation of the outliers to our advantage.


If $U$ itself were known, then at any given time, our problem would be similar to the dense error correction problem studied in \cite{error_correction_PCP_l1, Laska_exactsignal}. Of course the reason we need PCA is because $U$ is {\em unknown and cannot even be estimated.} Only, the PCs, $P_t:=(U)_{N_t}$ can be estimated.

Some recent work which actually is completely different from our current work, but may appear related (since it also uses CS and PCA or eigenvalue decomposition) includes \cite{fowler} and \cite{subspace_aug_music}.


\section{Notation}\label{sec_notions}
The set operations $\cup$, $\cap$ and $\setminus$ have the usual meanings. For any set $T \subset \{1 , \cdots n\}$, $T^c$ denotes its complement, i.e., $T^c : = \{i \in [1 , \cdots n]: i \notin T\}$, and $|T|$ denotes its cardinality, i.e., the number of elements in $T$. But $|a|$ where $a$ is a real number denotes the magnitude of $a$.

For a vector $v$, $v_i$ denotes the $i$th entry of $v$ and $v_T$ denotes a vector consisting of the entries of $v$ indexed by $T$. We use $\|v\|_p$ to denote the $\ell_p$ norm of $v$. The support of $v$, $\text{supp}(v)$, is the set of indices at which $v$ is nonzero, $\text{supp}(v): = \{i:\ v_i \neq 0\}$. We say that {\em $v$ is $s$-sparse} if $|\text{supp}(v)| \le s$. Sorting $|v_i|$ in descending order, we define the $p\%$-energy set of $v$ as $T_p:= \{|v_i| \geq \kappa\}$ where $\kappa$ is the largest value of $|v_i|$ such that $\|v_{T_p}\|_2^2 \geq p\% \|v\|_2^2$, i.e., $v_{T_p}$ contains the significantly nonzero elements of $v$.

For a matrix $A$, $A_i$ denotes the $i$th column of $A$ and $A_T$ denotes a matrix composed of the columns of $A$ indexed by $T$. We use $A_{T_1,T_2}$ to denote a submatrix of $A$ consisting of the rows indexed by $T_1$ and columns indexed by $T_2$. We use $A'$ to denote its transpose, and $A^{\dag}$ to denote its pseudoinverse. For a tall matrix $A$, $A^{\dag} = (A'A)^{-1}A'$. The Frobenius norm of matrix $A$ is denoted by $\|A\|_F$, i.e., $\|A\|_F : =\sqrt{ \Sigma_i \Sigma_j |A_{i,j}|^2}$.

For a tall matrix $P$, we use $\text{span}(P)$ to denote the subspace spanned by the column vectors of $P$.

For a diagonal matrix $D$, $D_T$ denotes a submatrix of $D$ consisting of the rows and columns indexed by $T$. In other words, $D_T$ is a diagonal matrix with $(D_T)_{j,j} = (D)_{T_j,T_j}$. Also, $\text{diag}(D)$ denotes the vector composed of the diagonal elements of $D$, i.e., $(\text{diag}(D))_i = D_{i,i}$.

We use $\emptyset$ to denote an empty set or an empty matrix.

\section{Recursive Projected Compressive Sensing} \label{rrpcp_sec}
We explain the Recursive Projected Compressive Sensing (ReProCS) algorithm below. In Sec. \ref{reprocs_assumption}, we discuss the implicit assumptions required for it.

\subsection{Recursive Projected Compressive Sensing (ReProCS) algorithm}
Let $\hat{P}_t$ be an estimate of the PC matrix $P_t$ at time $t$ and let $\hat{P}_{t,\perp}$ be \emph{an} orthogonal complement of $\hat{P}_t$. The column space of $\hat{P}_{t,\perp}$ is the null space of ${\hat{P}_t}'$, and hence $\hat{P}_{t,\perp}$ is not unique.
Using $\hat{P}_t$ and $\hat{P}_{t,\perp}$, we can rewrite $L_t = \hat{P}_t \alpha_t + \hat{P}_{t,\perp} \beta_t$ and hence $M_t$ as
\begin{equation}
M_t = \hat{P}_t \alpha_t +  \hat{P}_{t,\perp} \beta_t + S_t \nonumber
\end{equation}
where $\alpha_t := {\hat{P}_t}' L_t$ is the projection of $L_t$ into the subspace spanned by $\hat{P}_t$; and $\beta_t := (\hat{P}_{t, \perp})'L_t$ is the projection of $L_t$ into the subspace spanned by $\hat{P}_{t,\perp}$.

To approximately nullify the low rank part, $L_t$, we can project the data vector, $M_t$, into the space spanned by $\hat{P}_{t,\perp}$, i.e. compute
\begin{equation}
y_t:=A_t M_t, \ \ \text{where} \ \ A_t:= (\hat{P}_{t,\perp})'. \label{compute_yt}
\end{equation}
The dimension of the projected data vector, $y_t$, reduces to $n-r$ where $r:=\text{rank}(\hat{P}_t)$.
Notice that
\begin{equation}
y_t = A_t S_t + \beta_t, \  \text{where} \  \beta_t:= A_t L_t = A_t P_t (x_t)_{N_t}  \label{rrpcp_defn}
\end{equation}
If $\hat{P}_t \approx P_t$, then $A_t P_t \approx 0$, i.e. this nullifies most of the contribution of the low rank part, so that $\beta_t$ can be interpreted as small ``noise". Finding the $n$-dimensional sparse vector, $S_t$, from this $n-r$ dimensional projected data vector, $y_t$, now becomes the traditional noisy sparse reconstruction / compressive sensing (CS)  \cite{bpdn,candes,donoho} problem with the ``projected noise" $\beta_t$ resulting from the error in estimating $P_t$.
As long as $\|\beta_t \|_2$ is small and $A_t$ does not nullify any nonzero elements of $S_t$, we can recover $S_t$ by solving
\begin{equation}
\min_s \|s\|_1 \   \text{subject to}  \ \|y_t - A_t s\|_2 \le \epsilon \label{rpcp_cs}
\end{equation}
with $\epsilon$ chosen proportional to the ``noise" level, $\|\beta_t\|_2$. Denote its output by $\hat{S}_t$.
In practice, for large scale problems where $n$ is large, a less computationally and memory intensive way than (\ref{rpcp_cs}) is to solve
\begin{equation}
\min_s \|s\|_1 \   \text{subject to}  \ \|(I - \hat{P}_t \hat{P}_t')(M_t - s)\|_2 \le \epsilon \label{rpcp_cs_largescale}
\end{equation}
This is exactly equivalent to (\ref{rpcp_cs}) because $\hat{P}_{t,\perp}$ is an orthonormal complement of $\hat{P}_t$ satisfying $\hat{P}_t {\hat{P}_t}' + \hat{P}_{t,\perp} {\hat{P}_{t,\perp}}' = I$ and $\|{\hat{P}_{t,\perp}}' M_t\|_2 = \|\hat{P}_{t,\perp} {\hat{P}_{t,\perp}}' M_t\|_2$.

Using $\hat{S}_t$, we can then estimate
\begin{equation}
\hat{L}_t = M_t - \hat{S}_t \label{estimate_Lt}
\end{equation}
which can be used to recursively update the PC matrix estimate, $\hat{P}_t$, every-so-often to prevent the ``noise" $\beta_t$ from getting large (recursive PCA). We explain how to do this in Sec. \ref{sec_updatePt}.

In (\ref{rpcp_cs}), we need an appropriate parameter $\epsilon$ which should be proportional to the ``noise" term $\|\beta_t\|_2$. We set $\epsilon$ adaptively as $\epsilon = \|\hat{\beta}_{t-1}\|_2 = \|\hat{P}_{t, \perp}' \hat{L}_{t-1}\|_2$ at time $t$.

If the constraint in (\ref{rpcp_cs}) is too tight ($\epsilon$ is too small), it will give a solution with too many nonzero values. Moreover, as first explained in \cite{dantzig}, the solution of (\ref{rpcp_cs}) is always biased towards zero due to minimizing the $\ell_1$ norm. To address both these issues, as suggested in \cite{dantzig}, we can do support estimation followed by least squares (LS) estimation on the support, i.e., we can compute
\begin{eqnarray}
\hat{T}_t &=& \{i: \ (\hat{S}_t )_i \geq \gamma\} \label{supp} \\
(\hat{S}_t)_{\hat{T}_t} &=& ((A_t)_{\hat{T}_t})^{\dag} y_t, \
(\hat{S}_t)_{\hat{T}_t^{c}} = 0 \label{ls}
\end{eqnarray}
The low rank part, $L_t$, can then be estimated using this new $\hat{S}_t$ and (\ref{estimate_Lt}).

\begin{figure*}[t!]
\centerline{
\subfigure[Recursive Projected Compressive Sensing]{
\psfig{file = 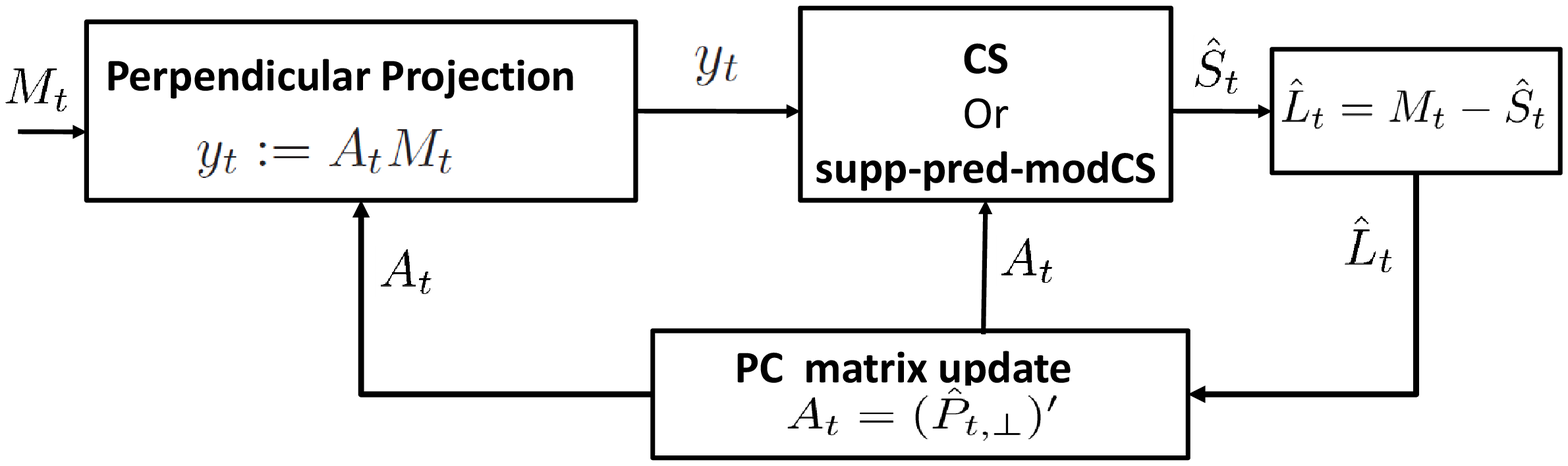, width = 7.5cm, height=3.5cm}\label{chart1}
}
\subfigure[support-predicted modified-CS]{
\psfig{file = 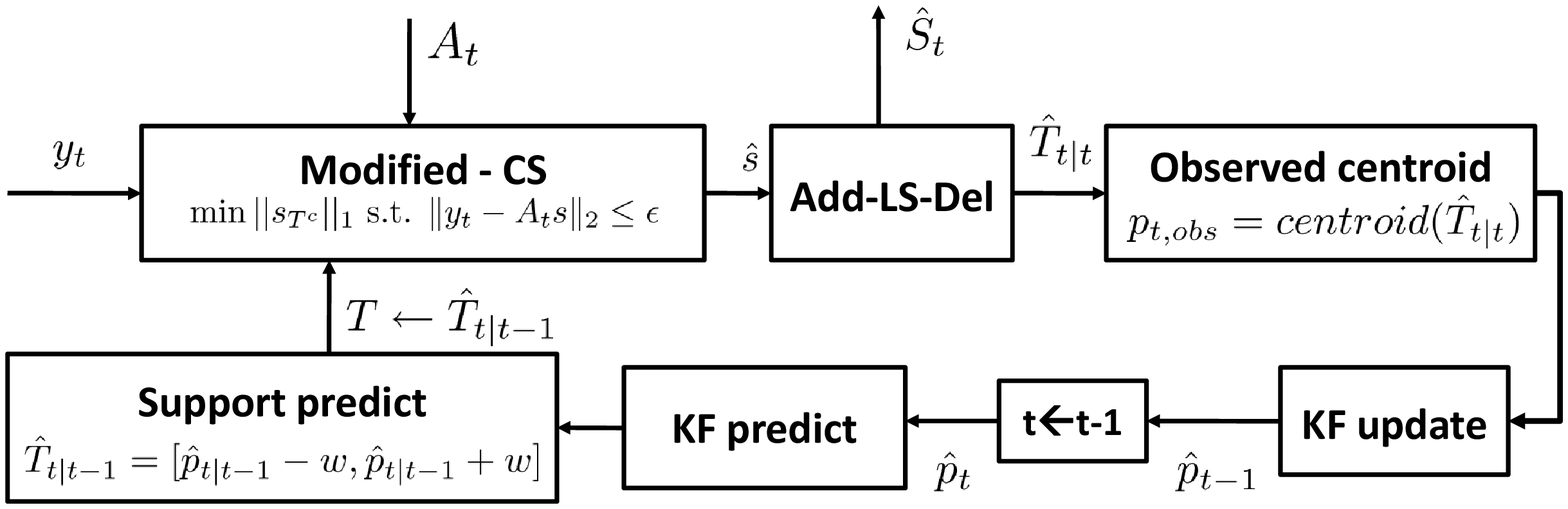, width = 10cm, height=4cm}\label{chart2}
}
}
\vspace{-0.1in}
\caption{Recursive Projected Compressive Sensing and support-predicted modified-CS}\label{complete_diag}
\vspace{-0.2in}
\end{figure*}

The block diagram of the above approach is shown in Fig. \ref{chart1} and the stepwise algorithm is given Algorithm \ref{algo_CS}.

\begin{algorithm}[t!]
\caption{ReProCS}\label{algo_CS}
At $t=t_0$, suppose a good estimate of PC matrix, $\hat{P}_{t_0}$ is available from training data.
For $t>t_0$, do the following:
\begin{itemize}
\item [1)] Let $A_t\leftarrow A_{t-1}$. Obtain $y_t$ by (\ref{compute_yt}).
\item [2)] Estimate $S_t$ by solving (\ref{rpcp_cs}) with $\epsilon = \|\hat{P}_{t, \perp}' \hat{L}_{t-1}\|_2$.
\item [3)] Support thresholding and least square estimation: do (\ref{supp}) and (\ref{ls}).
\item [4)] Estimate $\hat{L}_t = M_t - \hat{S}_t$.
\item [5)] Update $\hat{P}_t$ using Algorithm \ref{updatePt} which is based on \cite{sequentialSVD}. Update $A_t := (\hat{P}_{t,\perp})'$.
\item [6)] Increment $t$ by $1$ and go to step 1).
\end{itemize}
In practice, one can replace (\ref{rpcp_cs}) by (\ref{rpcp_cs_largescale}) with $\epsilon=\|(I - \hat{P}_t \hat{P}_t')\hat{L}_{t-1}\|_2$.
\end{algorithm}

\subsection{Implicit Requirements**}\label{reprocs_assumption} 


Suppose that $|T_t| \le s$ for all $t$, i.e. all $S_t$'s are $s$ sparse. Clearly, a necessary condition for the CS step of ReProCS, i.e. (\ref{rpcp_cs}), to recover the support of all $S_t$'s correctly is that $A_t=(\hat{P}_{t,\perp})'$ does not nullify any $s$-sparse vector, and thus does not nullify any nonzero part of any $S_t$. Or, equivalently, no $s$-sparse vector belongs to $\text{span}(\hat{P}_t)$. We can use the null space property (NSP) \cite{nsp} to show that a slightly stronger version of this requirement also serves as a sufficient condition, at least in the noise-free case.

Consider (\ref{rpcp_cs}). Assume that $\Phat_t = P_t$ so that the projected noise is indeed zero, i.e. $\beta_t = 0$. In this case, (\ref{rpcp_cs}) with $\eps=0$ will exactly recover any $s$-sparse $S_t$ if the following holds with a $\theta < 1$: $$\|(\eta)_T\|_1 \le \theta \|(\eta)_{T^c}\|_1$$ for all sets $T$ with $|T| \le s$ and for all $\eta \in \text{null}(A_t)$ \cite{nsp,wotao_yin}. Here $\text{null}(A_t):= \{\eta: A_t \eta = 0\}$ refers to the null space of $A_t$. In words, we need that all $s$-sparse or $s$-approximately-sparse vectors (i.e. vectors for which $\|(\eta)_{T^c}\|_1 \le \|(\eta)_T\|_1$ for some set $T$ with $|T| \le s$) do not lie in $\text{null}(A_t)$.  But $\text{null}(A_t) = \text{span}(\hat{P}_t)$. Thus, a sufficient condition for ReProCS to exactly recover $S_t$ in the noise-free case ($\beta_t=0$ and $\eps=0$ in (\ref{rpcp_cs})) is that no $s$-sparse or $s$-approximately-sparse vector lies in $\text{span}(\Phat_t)$.%

We expect that this should hold when the columns of $P_t$ are spread out enough (not sparse), which, in turn, should hold if the changes of the background, $L_t$, are  not localized in one or more small regions, e.g. due to water waves' motion in the video application. We show an example in Sec. \ref{expt_wave}. This observation will be analyzed in future work.

\begin{remark}
We should note that we can also get a sufficient condition using the restricted isometry property (RIP) \cite{decodinglp,candes_rip} and in fact it would hold even in the noisy case, but it is not as illustrative.
Let $\delta_s$ be the $s$-RIP constant  \cite{decodinglp} for the matrix $A_t$.  If the projected noise $\beta_t$ satisfies $\|\beta_t\|_2 \le \eps$ and if $\delta_{2s}< \sqrt{2}-1$, the reconstruction error is bounded by a constant times $\eps$, i.e. $\|\hat{S}_t - S_t\|_2 \le C(\delta_{2s}) \eps$ \cite{candes_rip}.
\end{remark}

\section{Support-predicted Modified-CS for ReProCS}
\label{supp_rrpcp} 
ReProCS needs to recover a $|T_t|$-sparse vector, $S_t$, from the projected data vector $y_t$ by solving a noisy CS problem (\ref{rpcp_cs}). The number of projected measurements available for the CS step (\ref{rpcp_cs}) is $n-r$, where $r=\text{rank}(\hat{P}_t)$. For a given $r$, if the support size of $S_t$, $|T_t|$, increases, or if $r$ increases for a given sparsity level, $|T_t|$, the ratio $\frac{|T_t|}{n-r}$ may become too large for CS to estimate $S_t$ accurately.
In this section we show how to utilize the correlated support changes of the $S_t$'s  to accurately recover them even when the ratio $\frac{|T_t|}{n-r}$ is too large for CS to work. We begin by first explaining the model on the support change of $S_t$. We then explain the {\em support-predicted modified-CS} algorithm, discuss why it is stable, and its extensions to more general cases. The complete ReProCS(modCS) algorithm is summarized in Algorithm \ref{algo_modCS}. 


\subsection{Model on the support change of $S_t$}\label{ModelS}
For explaining our ideas in a simple fashion, we consider a simple but realistic correlation model on $S_t$ inspired by the video application. We assume $S_t$ is a 1D foreground image with one moving foreground object that satisfies a constant velocity model described below. The extension to 2D images with multiple moving objects or to other correlation models is simple and is explained in Sec. \ref{extension_modCS}.

Let $p_t$ be the location of the foreground object's centroid at time $t$, let $v_t$ denote its velocity, and let $w$ denote its width.
Thus, its support is,
\begin{equation}
T_t = [p_t-w, p_t +w ]
\label{defTt}
\end{equation}
Let
\begin{equation*}
g_t := \left[ \begin{array}{c}p_t\\ v_t \\ \end{array} \right] \ \text{and}  \
G := \left[ \begin{array}{cc} 1 & \ 1 \\ 0 & \ 1 \\ \end{array}\right].
\end{equation*}
We assume the standard constant velocity model with small random acceleration on the object's motion \cite[Example V.B.2]{poor_book}, i.e.,
\begin{equation}
g_t = G g_{t-1} + \left[
                    \begin{array}{c}
                      0 \\
                      n_t \\
                    \end{array}
                  \right]  \label{motion_model}
\end{equation}
The modeling error (acceleration), $n_t$, is assumed to be bounded, with zero mean and variance $Q$.

\subsection{Support-predicted modified-CS (supp-pred-modCS)}\label{supppredmodcs}

The main idea of supp-pred-modCS is as follows. We use the above model in a Kalman filter (KF) to track the object's motion over time. The KF predicted location of the object and its size tells us its predicted support at the current time. This is then used to solve the modified-CS (modCS) problem and obtain an updated support estimate. The centroid (or median) of this support estimate tells us the observed location of the object, which may be erroneous because our support estimate is not perfect. This then serves as the noisy observation for the KF update step to update the current location and velocity estimates. We now explain each of these four steps.

 \ \\
\emph{Predict Location:}

Let $\hat{g}_{t|z} = [\hat{p}_{t|z} \ \hat{v}_{t|z} ]'$ denote the estimate of $g_t$ at time $t$ given measurements up to and including at time $z$. Similar rule applies for $\hat{T}_{t|z}$.
Let $\Sigma_{t|t-1}$, $\Sigma_{t|t}$ and $K_t$ denote the prediction and updated error covariance matrices and the Kalman gain used by the KF. Compute
\begin{eqnarray}
\hat{g}_{t|t-1} &=& G \  \hat{g}_{t-1|t-1} \label{KFpred1}, \ \text{and} \\
\Sigma_{t|t-1} &=& G \ \Sigma_{t-1|t-1} \ G' + \left[
                                                 \begin{array}{cc}
                                                   0 &  \\
                                                    & Q \\
                                                 \end{array}
                                               \right]
\label{KFpred2}
\end{eqnarray}

\begin{algorithm}[t!]
\caption{Support-predicted modified-CS}\label{algo_SuppPred}
\begin{itemize}
\item [1)] Predict centroid by (\ref{KFpred1}) and (\ref{KFpred2})
\item [2)] Predict support by (\ref{Tpred})
\item [3)] Update support
\begin{itemize}
\item Modified-CS: solve (\ref{rpcp_modCS_eq}) with $T = \hat{T}_t$ and $\epsilon = \|\hat{P}_{t, \perp}' \hat{L}_{t-1}\|_2$.
\item Add-LS-Del procedure:
\begin{eqnarray}
T_{\text{add}} &=& T \cup \{i \in T^c: |(\hat{s})_i| > \alpha_{\text{add}} \} \label{ald1} \\
(\hat{s})_{T_{\text{add}}} &=& ((A_t)_{T_{add}})^{\dag} y_t, \ (\hat{s})_{T_{\text{add}}^c} = 0 \label{ald2} \\
\hat{T}_{t|t} &=& T_{\text{add}} \setminus  \{i \in T_{\text{add}} : |(\hat{s})_i| < \alpha_{\text{del}} \} \label{ald3}\\
(\hat{S}_t)_{\hat{T}_{t|t}} &=& ((A_t)_{\hat{T}_{t|t}})^{\dag} y_t, \ (\hat{S}_t)_{\hat{T}_{t|t}^c} = 0 \label{ald4}
\end{eqnarray}
\end{itemize}
\item [4)] Update centroid by (\ref{KFupd1}), (\ref{KFupd2}) and (\ref{KFupd3}).
\end{itemize}
In practice, one can replace (\ref{rpcp_modCS_eq}) by
$\min_s \|s_{T^c}\|_1  \ \ \text{subject to} \ \ \|(I - \hat{P}_t \hat{P}_t')(M_t - s) \|_2 \le \epsilon$ with $\epsilon = \|(I - \hat{P}_t \hat{P}_t') \hat{L}_{t-1}\|_2$
\end{algorithm}

\ \\
\emph{Predict Support:}

Using the location prediction from above, compute the support prediction as
\begin{equation}
\hat{T}_{t|t-1} = [\hat{p}_{t|t-1} - w, \hat{p}_{t|t-1} + w] \label{Tpred}
\end{equation}

\ \\
\emph{Recover  $S_t$ and Update its Support using Modified-CS:}

Assuming $T = \hat{T}_{t|t-1}$ is a good support prediction, we can use it as the partial support knowledge for modCS which solves
\begin{equation}
\min_s \|s_{T^c}\|_1  \ \ \text{subject to} \ \ \|y_t  - A_t s \|_2 \le \epsilon
\label{rpcp_modCS_eq}
\end{equation}
i.e., it tries to find the solution that is sparsest outside the set $T$ among all solutions that satisfy that data constraint.
Let $\hat{s}$ be the solution of (\ref{rpcp_modCS_eq}) with $T = \hat{T}_{t|t-1}$.
As explained in \cite{stability_allerton}, $\hat{s}$ is biased towards zero along $T^c$ (because of the $\ell_1$ term) and it may be biased away from zero along $T$ (there is no cost on $s_T$ and the only constraint is the data constraint). The elements which are missing from the support prediction, $\Delta_{t|t-1}:= T_t \setminus T$ are a subset of $T^c$ where as the extra elements in the support prediction, $\Delta_{e,t|t-1}: = T \setminus T_t$ are a subset of $T$. If we use a single threshold for support estimation as in (\ref{supp}), we will need a small threshold to ensure that most elements of $\Delta_{t|t-1}$ get correctly detected, but a large one will be needed to ensure that elements of $\Delta_{e,t|t-1}$ are deleted. Thus for a given threshold, one or the other cannot be done well.

A better approach is to use the Add-LS-Del procedure summarized in step 3 of Algorithm \ref{algo_SuppPred}. This was first introduced in our older work \cite{LS-CS-residual,kfcsicip} and simultaneously also in \cite{SubSpaceCS,cosamp}. 
The addition step threshold, $\alpha_{\text{add}}$, needs to be just large enough to ensure that the matrix used for LS estimation, $A_{T_{\text{add}}}$ is well-conditioned. If $\alpha_{\text{add}}$ is chosen properly, the LS estimate on $T_{\text{add}}$ will have smaller error than the modCS output. As a result, deletion will be more accurate when done using this estimate. This also means that one can use a larger $\alpha_{\text{del}}$ to ensure quicker deletion of extras. We denote the final support estimate output of the Add-LS-Del procedure by $\hat{T}_{t|t}$.


\ \\
\emph{Update Location:}

The centroid of the updated support estimate, $\hat{T}_{t|t}$, obtained above, serves as the ``observed" location for the KF, $p_{t,\text{obs}}$, i.e.
\begin{equation}
p_{t,\text{obs}} = \text{centroid}(\hat{T}_{t|t}) := \frac{1}{|\hat{T}_{t|t}|} \sum_{i \in \hat{T}_{t|t}} i \label{supportmean}
\end{equation}
The following is a valid model for $p_{t,\text{obs}}$
\begin{eqnarray}
p_{t,\text{obs}} =  H g_t + \omega_t, \ H := [1 \ 0]   \label{obsmod4}
\end{eqnarray}
where $\omega_t$ is the observation error, which is bounded, zero mean and has a variance $R$. The observation error arises because there are extras and misses in $\hat{T}_{t|t}$ and hence $p_{t,\text{obs}} = \text{centroid}(\hat{T}_{t|t}) \neq \text{centroid}(T_t) = p_t$. 
In practice, using the median instead of centroid provides a better observed estimate especially when there are many extras in the support estimate. We discuss how to set $R$ in Sec. \ref{extension_modCS}.

The KF update step is as follows \cite[Chapter 5]{poor_book}.
\begin{eqnarray}
K_t &=& \Sigma_{t|t-1} \ H' \ (H\ \Sigma_{t|t-1} \ H'+R)^{-1} \label{KFupd1} \\
\hat{g}_{t|t} &=& \hat{g}_{t|t-1} + K_t \ (p_{t,\text{obs}}-H\ \hat{g}_{t|t-1}) \label{KFupd2}\\
\Sigma_{t|t} &=& \Sigma_{t|t-1} - K_t \ H \ \Sigma_{t|t-1} \label{KFupd3}
\end{eqnarray}

The above algorithm is summarized in Algorithm \ref{algo_SuppPred} and also in the block diagram of Fig.\ref{chart2}. A complete algorithm incorporating the idea of supp-pred-modCS into ReProCS is given in Algorithm \ref{algo_modCS}.

\begin{algorithm}[t!]
\caption{ReProCS(modCS)}
\label{algo_modCS}
At $t=t_0$, suppose a good estimate of PC matrix, $\hat{P}_{t_0}$ is available from training data.
For $t>t_0$, do the following:
\begin{itemize}
\item [1)] Let $A_t\leftarrow A_{t-1}$. Obtain $y_t$ by (\ref{compute_yt}).
\item [2)] Estimate $S_t$ by Algorithm \ref{algo_SuppPred}.
\item [3)] Estimate $\hat{L}_t = M_t - \hat{S}_t$.
\item [4)] Update $\hat{P}_t$ using Algorithm \ref{updatePt} which is based on \cite{sequentialSVD}. Update $A_t := (\hat{P}_{t,\perp})'$.
\item [5)] Increment $t$ by $1$ and go to step 1).
\end{itemize}
\end{algorithm}

\subsection{Stability: Main Ideas**}\label{sec_stability}

The ReProCS algorithm is a recursive approach. In ReProCS(modCS), the sparse recovery algorithm, support-predicted modified-CS, is itself recursive. Hence an important question is when and why will it be stable (error bounded by a time-invariant and small value)? Our simulations given in Sec. \ref{results} do indicate that it is stable.  In this section, we intended to provide the main arguments of why this can be shown analytically. Due to lack of space, we have moved this discussion to Supplementary Material. Based on reviewer comments, it can later be incorporated into the main paper by shortening other sections.

\subsection{Discussion and Extensions**} \label{extension_modCS}

It is easy to see that, if $p_t$ is the centroid of $\hat{T}_{t|t}$, then the observation error can be bounded as $|\omega_t| \le  |\Delta_{t|t}| \frac{w}{2w+1 - |\Delta_{t|t}|}  + |\Delta_{e,t|t}| \frac{\max_{j \in \Delta_{e,t|t}} |j-p_t|}{2w}$. We show this in the stability discussion (see Supplementary Material). Denote this bound by $B$. A possible way to set the variance, $R$, of the observation error $\omega_t$ is to assume that it has the maximum variance distribution for a given bound (i.e. is uniformly distributed). Using this, $R = B^2/3$. But doing this will require roughly knowing the final number of misses and extras. This will depend on the intensity distribution of the moving object and of the background. It can be estimated only if a training sequence with the same object is available. But even without this, the following qualitative fact always holds.  If the background and foreground values are quite different, the updated support estimate will be more accurate. So, clearly $\omega_t$ will be small. This can also be seen from the bound on $|\omega_t|$ which is directly proportional to the updated support errors. In this case, $R$ should be smaller.
In general, $R$ should be of the same order as $Q$ if we want the KF to compute a weighted average of the new observed location and the predicted one. The smaller (larger) the value of $R$ compared to $Q$, the greater (lesser) the KF depends on the observed location (centroid/median of the updated support).%



For the sake of simplicity, we presented our idea of supp-pred-modCS for a 1D image sequence using a very simple correlation model on the support change of the sparse part $S_t$. The sparse part was assumed to contain a single translating object. However, we can extend it to other more general cases. The extension to 2D image sequences is easy. Position and velocity will each be a two dimensional vector. The support prediction and the location update steps will need simple changes to handle 2D motion. The extension to multiple objects is also easy if they have sufficiently different intensities; if their supports do not overlap for too long; and if their motion is independent. One can have a separate KF for tracking the motion of each object. To obtain the observed locations of the different objects one can use intensity thresholds to obtain their respective supports. This is done in our experiment of Sec. \ref{expt_modcs}.

The case of multiple and changing number of objects; objects that can enter and leave the scene; and objects that cannot be separated by intensity thresholding but need more sophisticated segmentation techniques will be studied in future work. Also, in the current work we only assume translating objects. The extension to any other linear model (e.g. affine deformation model in case of video or any other linear model in other applications) is easy.
Finally, so far we have not used the fact that the support will consist of one or a few contiguous blocks in the support update step. Using this can significantly help remove arbitrary extras. One way to use this fact is to use the model-based CS idea of \cite{modelcs}.

\section{Recursive PCA**} \label{sec_updatePt}

As explained in Sec. \ref{intro_prob}, $L_t = U x_t$ where $U$ is an unknown orthonormal matrix and $x_t$ is a sparse vector. The support set of $x_t$, $N_t$, changes every-so-often and the nonzero elements of $x_t$ are spatially uncorrelated. Thus, $P_t$ also changes every-so-often. One realistic generative model on $x_t$ and hence on $L_t$ is given in the Appendix. When new PCs appear, we need to detect them timely before the projected noise, $\beta_t$, seen by CS gets too large. When some old PCs vanish, we also need to remove them from $\hat{P}_t$. Otherwise, $r=\text{rank}(\hat{P}_t)$ will keep increasing and therefore the number of projected measurements, $n-r$, will keep decreasing and may become too small for CS or modCS to work.


At the initial time, $t=t_0$, if the training sequence, $[M_1, \cdots, M_{t_0}]$, does not contain any sparse part, we let $[\hat{L}_1, \cdots, \hat{L}_{t_0}] = [M_1, \cdots, M_{t_0}]$.  Usually such a sequence is easy to obtain, e.g. in video this means having a sequence with no foreground moving objects. If $[M_1, \cdots, M_{t_0}]$ does contain the sparse part, but its support is small enough or uncorrelated enough so that PCP works, then $[\hat{L}_1, \cdots, \hat{L}_{t_0}]$ can be obtained by solving PCP, i.e. (\ref{PCP}), with $M = [M_1, \cdots, M_{t_0}]$. 

The initial PC matrix, $\hat{P}_{t_0}$, is estimated by computing a singular value decomposition (SVD) of $[\hat{L}_1, \cdots, \hat{L}_{t_0}] $, and retaining singular values above a threshold $\alpha_0$, i.e., we compute $[\hat{L}_1, \cdots, \hat{L}_{t_0}] \overset{SVD}{=} P \Lambda V'$; set $T \leftarrow \{i: (\Lambda)_{i,i} > \alpha_0 \}$ and set  $\hat{P}_{t_0}  \leftarrow P_{T}$, $\hat{\Lambda}_{t_0} \leftarrow \Lambda_{T,T}$, $\hat{V}_{t_0} \leftarrow V_T$.
Thus, all but the singular values above $\alpha_0$ are zeroed to give a truncated SVD that approximates the data. 
Usually, $\alpha_0$ is picked according to the distribution of the singular values. 
If the matrix  $[\hat{L}_1, \cdots, \hat{L}_{t_0}] $ is exactly low rank, $\alpha_0$ can zero. Otherwise it can be picked so that $T$ is the $p$\%-energy set of $\text{diag}(\Lambda)$ for $p$ large enough.%

{\em We use $\hat{V}_t$ to only explain how we update $\hat{P}_t$, but we do not need to compute and store $\hat{V}_{t}$}.

\begin{algorithm}[]
\caption{Recursive PCA}\label{updatePt}
\textbf{Initialization}: 
Compute $[\hat{L}_1, \cdots, \hat{L}_{t_0}] \overset{SVD}{=} P \Lambda V'$; set $T \leftarrow \{i: (\Lambda)_{i,i} > \alpha_0 \}$,  $\hat{P}_{t_0}  \leftarrow P_{T}$, $\hat{\Lambda}_{t_0} \leftarrow \Lambda_{T,T}$. 

 Let $D=\emptyset$. At each time $t > t_0$, store $\hat{L}_t$ in $D$, i.e., $D \leftarrow [D \ \hat{L}_t]$. Do the following:

\begin{itemize}
\item [1)] If there are less than $\tau$ frames in $D$,
\begin{itemize}
\item [-] Let $\hat{P}_t \leftarrow  \hat{P}_{t-1}$, $\hat{\Lambda}_t \leftarrow \hat{\Lambda}_{t-1}$.
\end{itemize}

\item[2)] If there are $\tau$ frames in $D$
\begin{itemize}

\item[2a)] Remove decayed directions from $\hat{P}_{t-1}$
\begin{eqnarray*}
\hat{\Sigma}_{\mathcal{E}} &=& \frac{1}{\tau} {\hat{P}_{t-1}}'DD'\hat{P}_{t-1} \\
\hat{N}_{\mathcal{E}} &=& \{ i: (\hat{\Sigma}_{\mathcal{E}})_{i,i} \geq \alpha \} \\
\hat{P}_{t-1} &\leftarrow& (\hat{P}_{t-1})_{\hat{N}_{\mathcal{E}} }, \ \hat{\Lambda}_{t-1} \leftarrow  (\hat{\Lambda}_{t-1})_{\hat{N}_{\mathcal{E}},\hat{N}_{\mathcal{E}}} \\
\hat{r}_{t-1} &\leftarrow& \text{rank}(\hat{P}_{t-1})
\end{eqnarray*}

\item[2b)] Update $\hat{P}_t$ and $\hat{\Lambda}_t$ by incremental SVD \cite{sequentialSVD}
\begin{itemize}
\item[-] Compute $C = \hat{P}_{t-1}'D$ and $E = D-\hat{P}_{t-1} C$
\item[-] Compute QR decomposition of $E$: $E  \overset{QR}{=} JK$
\item[-] Compute SVD of $\left[ \begin{array}{cc} \hat{\Lambda}_{t-1} & C \\ 0 & K \\\end{array} \right]$ :$\left[ \begin{array}{cc}\hat{\Lambda}_{t-1} & C \\ 0  & K \\\end{array}\right] \overset{SVD}{=} P_r \Lambda_rV_r' $
\item[-] Update $\hat{P}_t$ and $\hat{\Lambda}_t$  as
\begin{equation*}
\hat{P}_{t} \leftarrow [\hat{P}_{t-1} \ \ J ] P_r , \ \ \hat{\Lambda}_t \leftarrow \Lambda_r
\end{equation*}
\end{itemize}

\item [2c)] Add new PCs with large variance
\begin{eqnarray*}
\hat{N}_{t} &=& \{ 1,\cdots, \hat{r}_{t-1}\} \cup \{ i> \hat{r}_{t-1} : \frac{(\hat{\Lambda}_{t})_{i,i}^2}{\tau} \geq \alpha\} \\
\hat{P}_{t} &\leftarrow& (\hat{P}_{t})_{\hat{N}_{t} } , \ \hat{\Lambda}_{t} \leftarrow (\hat{\Lambda}_{t})_{\hat{N}_{t} }
\end{eqnarray*}

\item[2e)] Reset $D=\emptyset$.
\end{itemize}

\end{itemize}

\end{algorithm}


For $t>t_0$, we obtain $\hat{L}_t$ by ReProCS. In the recursive PCA step of ReProCS, we update $\hat{P}_t$ every $\tau$ frames. Alternatively, we can also do this whenever $\eps = \|(I- \Phat_{t-1} \Phat_{t-1}') \hat{L}_{t-1}\|_2$ exceeds a threshold. 

The PC update is done as follows. The complete algorithm is summarized in Algorithm \ref{updatePt}.
At $t=t_0 + k \tau$, let $ D = [\hat{L}_{t-\tau+1}, \cdots, \hat{L}_t]$.
In step 2a), we first estimate the variance of $D$ along the columns of $\hat{P}_{t-1}$ and remove the PCs along which the variance is below a threshold $\alpha$. Let $\sigma_{\text{min}}$ be the smallest nonzero singular value of $\hat{\Lambda}_{t_0}$. 
We use $\alpha = 0.5 \sigma_{\text{min}}^2$. Note that after doing step 2a), the column vectors of $\hat{P}_{t-1}$ contain all the non-decayed PCs.


In step 2b), we rotate $\hat{P}_{t-1}$ and find the new PCs based on the idea of incremental SVD \cite{sequentialSVD}. We first decompose the new collected data $D$ into two components $C$ and $E$, where $C = {\hat{P}_{t-1}}'D$ and $E = D-\hat{P}_{t-1} C$. The parallel component $C$ rotates the existing singular vectors and the orthogonal component $E$ estimates the new PCs \cite{sequentialSVD}. Let $E \overset{QR}{=} JK$ be a QR decomposition of $E$. Notice that $\hat{P}_{t-1}$ and $J$ are orthogonal, i.e., $\hat{P}_{t-1}'J=0$. The column vectors of $J$ are the basis vectors of the subspace spanned by the new PCs.
It is easy to see that \cite{sequentialSVD}
\begin{equation*}
[\hat{P}_{t-1} \hat{\Lambda}_{t-1} {\hat{V}_{t-1}}' \ \  D] = [\hat{P}_{t-1} \ \ J] \
\left[ \begin{array}{cc} \hat{\Lambda}_{t-1} & C \\  0 & K \\ \end{array} \right] \
\left[\begin{array}{cc}\hat{V}_{t-1} & 0 \\0 & I \\ \end{array} \right]'
\end{equation*}
Let
$\left[ \begin{array}{cc} \hat{\Lambda}_{t-1} & C \\  0 & K \\ \end{array} \right] \overset{SVD}{=} P_r \Lambda_r V_r'$.
Clearly,
\begin{equation*}
[\hat{P}_{t-1} \hat{\Lambda}_{t-1} {\hat{V}_{t-1}}' \ \ D] \overset{SVD}{=} ([\hat{P}_{t-1} \ \ J ] P_r) \Lambda_r (\left[ \begin{array}{cc} \hat{V}_{t-1} & 0 \\ 0 & I \\ \end{array} \right] V_r)'
\end{equation*}
In words, $P_r$ rotates the old PCs, $\hat{P}_{t-1}$, and the new basis vectors, $J$, to the current PCs, i.e., $\hat{P}_t = [\hat{P}_{t-1} \ \ J ] P_r$. Also, the singular values along $\hat{P}_t$ are the diagonal elements of $\hat{\Lambda}_t = \Lambda_r $.

Let $\hat{r}_{t-1} = \text{rank}(\hat{P}_{t-1})$ and let $\hat{N}_{\mathcal{A}} = \{\hat{r}_{t-1}+1,\cdots,\text{rank}(\hat{P}_t)\}$. 
If $\hat{P}_{t-1} \approx P_{t-1}$, the old PCs are already correctly estimated and do not need to be rotated. Under this assumption, $(\hat{P}_t)_{\hat{N}_{\mathcal{A}}}$ contains the new PCs. The variance of $D$ along the columns of $(\hat{P}_t)_{\hat{N}_{\mathcal{A}}}$ is given by the diagonal elements of  $\frac{1}{\tau}(\hat{\Lambda}_t)_{\hat{N}_{\mathcal{A}}}^2$. Therefore, to only retain the new PCs with large variance, we threshold on $\frac{1}{\tau}(\hat{\Lambda}_t)_{\hat{N}_{\mathcal{A}}}^2$ in step 2c).

Recall that $n - r$ (where $r=\text{rank}(\hat{P}_t)$) is the number of projected measurements and $\beta_t$ is the noise seen by CS or modified-CS (see (\ref{rrpcp_defn})). By choosing different values of the threshold, $\alpha$, there is a tradeoff between making $\beta_t$ small and keeping $n- r$ large. A smaller $\alpha$ means we retain more directions in $\Phat_t$. This means that $n-r$ is smaller (fewer measurements) but the projected noise, $\beta_t = (I - \Phat_t \Phat_t') L_t$ is also smaller.  When the nonzero elements of the sparse part, $S_t$, have very small magnitudes, we need a smaller $\alpha$ to ensure that the noise, $\beta_t$, is sufficiently small compared to $S_t$. 

\section{Experimental Results} \label{results}
We compare the performance of ReProCS and ReProCS(modCS)\footnote{we use YALL1 $\ell_1$ minimization toolbox \cite{yall1} to solve (\ref{rpcp_cs}) and (\ref{rpcp_modCS_eq}). Its code is available at \url{http://yall1.blogs.rice.edu/}.} with two recursive robust PCA methods -- incremental robust subspace learning (iRSL) \cite{Li03anintegrated}, adapted (outlier-detection enabled) incremental SVD (adapted-iSVD) \cite{sequentialSVD}, and two batch robust PCA methods -- principal components' pursuit (PCP)\footnote{We use Accelerated Proximal Gradient algorithm\cite{apg} and Inexact ALM algorithm \cite{alm} (designed for large scale problems) to solve PCP (\ref{PCP}). The code is available at \url{http://perception.csl.uiuc.edu/matrix-rank/sample_code.html}.} \cite{rpca}, robust subspace learning (RSL)\footnote{The code of RSL is available at http://www.salleurl.edu/~ftorre/papers/rpca/rpca.zip.} \cite{Torre03aframework}. We also show a comparison with  simple thresholding based background subtraction (BS) for the last experiment.

In Sec. \ref{expt_n100} and Table \ref{tabel_intro}, we compare ReProCS with  iRSL\cite{Li03anintegrated}, adapted-iSVD\cite{sequentialSVD}, PCP \cite{rpca} and RSL \cite{Torre03aframework}. As can be seen, iRSL and adapted-iSVD fail in case of both small magnitude $S_t$'s and large support-sized $S_t$'s. RSL can handle large support size, but not small magnitude $S_t$'s while the opposite is true for PCP. In later comparisons we only compare against PCP and RSL.
%
%
In Sec. \ref{expt_comparison}, Table \ref{table1} and Fig. \ref{expt1}, we give a detailed comparison of ReProCS and PCP. PCP has large error whenever the support sets of the $S_t$'s are correlated and their sizes are large. Comparison with RSL is also shown.

In Sec. \ref{expt_modcs} and Fig. \ref{expt3_two}, we show an example where ReProCS does not work and ReProCS(modCS) is needed. Here the support size of $S_t$, $|T_t| \approx 0.51n$ while the number of projected measurements is only about $n-r \approx 0.8n$.

In Sec. \ref{expt_wave}, we show comparisons on a partly-real video sequence where we overlay simulated sparse foreground images on real background images. The background sequence has nonzero mean. The background variations, e.g. those due to water waves, lead to the low-rank part having covariance matrix rank as large as $20\%$ of image size. Because the variations due to the water waves are not localized in one or more small image regions, the resulting projection matrix, $A_t = (\hat{P}_{t,\perp})'$, is ``incoherent" enough compared to the sparse foreground image and so CS and modCS work. 

{\em All our code will be posted on the first author's webpage, \url{http://www.ece.iastate.edu/~chenlu}. Videos of all the above experiments and some results on fully real video sequences are also posted here.}

\subsection{Comparison of ReProCS with adapted-iSVD, iRSL, RSL and PCP} \label{expt_n100}

In this experiment, the measurement at time $t$, $M_t := L_t +S_t$, is an $n \times 1$ vector with $n=100$. The low rank part, $L_t = U x_t$, where $U$ is an $n \times n$ orthonormal matrix  (generated by first generating an $n \times n$ matrix with entries randomly sampled from a standard Gaussian distribution, and then orthonormalizing it using the Gram-Schimidt process) and $x_t$ is an $n\times 1$ sparse vector with support set $N_t$ with size $|N_t| \approx 0.2n$. Thus the PC matrix at time $t$ is $P_t := (U)_{N_t}$. The sparse vector, $x_t$, is simulated using the regression model described in the Appendix with $f=0.5$, $f_d = 0.1$, and $\theta=0.5$. Initially there are $20$ PCs with variances $10^4, 0.7079\times10^4, 0.7079^2 \times 10^4, \cdots, 14.13$. At $t=t_0+5$, two new PCs with variance $50$ and $60$ enter the PC basis, $P_t$, and the values of $x_t$ along two old PCs start to decay exponential to zero. The sparse part, $S_t$, is an $n\times 1$ vector with its support set denoted by $T_t$.
%
%
For $1 \le t \le t_0=2000$, $S_t=0$ and hence $M_t=L_t$. For $t>t_0$, the support set of $S_t$, $T_t$, is generated in a correlated fashion: $S_t$ contains either one or four nonzero strips (small and large support size cases). Each strip has $9$ nonzero elements and it can either stay static with a large probability $0.8$ or move top/down with a small probability $0.1$ independently of the others. Thus the support sets are both spatially and temporally highly correlated. The magnitude of the nonzero elements of $S_t$ is fixed at either 100 (large) or 10 (small). The two cases are plotted in Fig. \ref{expt1_intro}.

We tabulate our results in Table \ref{tabel_intro}.  The first $t_0$ frames can be used to obtain the initial PC matrix, $\Phat_{t_0}$, by standard PCA on $[M_1, \cdots, M_{t_0}] = [L_1, \cdots, L_{t_0}]$. The same $\Phat_{t_0}$ is used by ReProCS (Algorithm \ref{algo_CS}) and by the two other recursive algorithms -- iRSL \cite{Li03anintegrated} and adapted-iSVD \cite{sequentialSVD}. For ReProCS, the support estimation threshold in (\ref{supp}), $\gamma$, can be set a little lower than the minimum nonzero magnitude that we would like to correctly detect. Thus we set $\gamma =  a \min_{i \in T_t} |(S_t)_i|$ for an $a<1$. We used $a = 0.2$ in the large magnitude $S_t$ case, but $a=0.3$ in the small magnitude $S_t$ case (using a larger $a$ in this case ensures that $\gamma$ is greater than the noise level). We update $\hat{P}_t$ for every $\tau = 20$ frames by Algorithm \ref{updatePt} with $\alpha=5$.
iRSL \cite{Li03anintegrated} solves the recursive robust PCA problem by weighting each data point according to its reliability (thus soft-detecting and down-weighting likely outliers). If the final goal is outlier detection (e.g. detecting foreground moving objects), this is done in \cite{Li03anintegrated} by thresholding on the difference between the current data vector and its projection into the PCA space, i.e. by thresholding on $(I - \Phat_t \Phat_t') M_t$. We also compare against an adapted version of \cite{sequentialSVD} (we call it adapted-iSVD). We provide adapted-iSVD the outlier locations also by thresholding on  $(I - \Phat_t \Phat_t') M_t$.  It fills in the corrupted locations of $L_t$ by imposing  that $L_t$ lies in $\text{span}(\Phat_t)$. We used a threshold of $0.5 \min_{i \in T_t} |(S_t)_i|$ for both iRSL and adapted-iSVD (we also tried various other options for thresholds but with similar results).
PCP \cite{rpca} is a batch method that needs to wait until all data frames $M = [M_1,\cdots, M_t]$ are available and then estimates the low rank part $L = [L_1,\cdots, L_t]$ and the sparse part $S=[S_1,\cdots, S_t]$ simultaneously by solving the convex optimization problem of (\ref{PCP}) with $\lambda = 1/ \sqrt{\text{max}(n,t)}$ as suggested in \cite{rpca}. RSL \cite{Torre03aframework} is another state-of-art batch robust PCA method that aims at recovering a good low-rank approximation that best fits the majority of the data. RSL solves a nonconvex optimization via alternative minimization based on the idea of soft-detecting and down-weighting the outliers. For RSL and PCP, $\hat{S}_t$ is not exactly sparse and therefore we estimate the support, $\hat{T}_t$, by thresholding on $\hat{S}_t$ with a threshold picked in the same way for adapted-iSVD and iRSL.


\begin{table*}
\centerline{
\subtable[$(S_t)_i = 100, \forall i \in T_t$ and $(S_t)_i = 0, \forall i \in T_t^c$]{
    \begin{tabular}{ | l | c | c | c | c | c |}
    \hline
    $|T_t|/n$   & ReProCS  & adapted-iSVD & iRSL  & RSL  & PCP   \\
    \hline
    $9\%$  & $0.00002$ & $0.0381$ & $0.9817$ & $0.0001$ & $0$\\   
    \hline
    $36\%$ & $0.0119$  & $0.3172$ & $0.9813$ & $0.0445$ & $0.5553$\\
    \hline
    \end{tabular}
}
\subtable[$(S_t)_i = 10, \forall i \in T_t$ and $(S_t)_i = 0, \forall i \in T_t^c$]{
    \begin{tabular}{ | l | c | c | c | c | c |}
    \hline
    $|T_t|/n$   & ReProCS  & adapted-iSVD & iRSL  & RSL  & PCP  \\
    \hline
    $9\%$  & $0.0480$ & $0.1949$ & $0.9890$ & $4.9618$ & $0.0946$ \\
    \hline
    $36\%$  & $0.0563$ & $0.4043$ & $0.9920$ & $1.2482$ & $0.5786$ \\
    \hline
    \end{tabular}
}
}
\vspace{-0.1in}
\caption{\small{Normalized MSE $\mathbb{E}\|O-\hat{O}\|_F^2 / \mathbb{E} \|O\|_F^2$ of various methods. The experiment is described in Sec. \ref{expt_n100}. Here, $|T_t|/n$ is the sparsity ratio of $S_t$.}}\label{tabel_intro}
\vspace{-0.15in}
\end{table*}

iRSL was designed for a video application and hence it directly detects the foreground overlay's support by thresholding on  $(I-\hat{P}_{t}\hat{P}_{t}')M_t$ (it never computes $S_t$). Thus, in this experiment, for all algorithms, we compare the reconstruction error of the foreground overlay, $(O_t)_{T_t} = (M_t)_{T_t}$ and $(O_t)_{T_t^c} = 0$, where $T_t$ is the support set of $S_t$. 
%
In Table \ref{tabel_intro}, we compare the normalized mean square error (MSE), $\mathbb{E}\|O-\hat{O}\|_F^2 / \mathbb{E} \|O\|_F^2$ where $O:=[O_{t_0+1},\cdots,O_{t_0+100}]$.
The expected value is computed using $100$ times Monte Carlo averaging.

As can be seen, ReProCS gives small error in all four cases. ReProCS is able to successfully recover both small magnitude and fairly large support-sized $S_t$'s because it operates by first approximately nullifying $L_t$, i.e. computing $y_t: = (I-\hat{P}_{t}\hat{P}_{t}') M_t$, and then recovering $S_t$ by solving (\ref{rpcp_cs}) that enforces the sparsity of $S_t$.  Adapted-iSVD and iRSL also approximately nullify $L_t$ by computing the same $y_t$, but they directly use $y_t$ to detect or soft-detect (and down-weight) the support of $S_t$. Using (\ref{eq1}), $y_t$ can be rewritten as $y_t = S_t + (-\hat{P}_{t}\hat{P}_{t}'S_t) + \beta_t$ where $ \beta_t:=(I-\hat{P}_{t}\hat{P}_{t}')L_t$. As the support of $S_t$ increases, the interference due to $(-\hat{P}_{t}\hat{P}_{t}'S_t)$ becomes large, resulting in wrong estimates of $S_t$. Since adapted-iSVD and iRSL are recursive methods, this, in turn, results in wrong PC matrix updates, thus also causing $\beta_t$ to become large and finally causing the error to blow up.  Using $y_t$ directly to detect the support of $S_t$ also becomes difficult when the magnitude of the nonzero $S_t$'s is small compared to $\beta_t$ and this is another situation where adapted-iSVD and iRSL fail. RSL is able to handle larger support size of $S_t$'s because it uses the entire sequence of $t_0+100$ $M_t$'s jointly and the first $t_0=2000$ $M_t$'s have $S_t=0$. But when the magnitude of the nonzero $S_t$'s is small, RSL also fails. PCP is able to always deal with small support sized $S_t$'s (even when their magnitude is small). 
But it fails for large support sizes of $S_t$'s, particularly since they are also correlated. PCP is much more robust to independently generated $S_t$'s with similar ratios of $|T_t|/(n-r)$ (shown next).

\subsection{Recovery of random or structured sparse outliers}\label{expt_comparison} 

In this experiment, we consider the problem of tracking moving blocks in a simulated image sequence of size $32 \times 32 \times (t_0+200)$. Thus, the data vector at time $t$, $M_t$, is $n = 32^2 = 1024$ dimensional. Note that $M_t = L_t + S_t$ with $L_t$ and $S_t$ generated as described below.
We let $L_t = U x_t$ where $U$ is an orthonormal $n \times n$ matrix generated as before and $x_t$ is an $n$ dimensional sparse vector.
The sparse vector, $x_t$, is simulated using the regression model described in the Appendix with $f=0.5$, $f_d = 0.1$, and $\theta=0.5$. In Table \ref{table1}\ref{table_random_32} and Table \ref{table1}\ref{table_structured_32}, initially there are $32$ PCs with variances $10^4, 0.8058\times10^4, 0.8058^2 \times 10^4, \cdots, 12.4$.
In Table\ref{table1}\ref{table_random_128} and Table\ref{table1}\ref{table_structured_128}, initially there are $128$ PCs with variances $10^4, 0.9471\times 10^4, 0.9471^2 \times 10^4, \cdots, 10$.
In all four cases, at time $t = t_0 + 5$, two new PCs with variance $50$ and $60$ enter the PCs basis and the value of $x_t$ along two old PCs begins to decay exponentially to zero. At time $t = t_0 +50$, another two new PCs with variance $55$ and $65$ enter the PCs basis and the value of $x_t$ along two old PCs begins to decay exponentially to zero.
For $1 \le t \le t_0$, $S_t=0$ and hence $M_t=L_t$. For $t>t_0$, $S_t$ has nonzero elements. Each nonzero element of $S_t$ has constant value $5$, which is much smaller than those of $L_t$'s. We generate the support of $S_t$, $T_t$, in the following two different ways:
\begin{itemize}
\item In Table \ref{table1}\ref{table_random_32} and Table \ref{table1}\ref{table_random_128}, for a given support size $|T_t|$, $T_t$ is generated uniformly at random.
\item In Table \ref{table1}\ref{table_structured_32} and Table \ref{table1}\ref{table_structured_128}, $T_t$ is spatially and temporal correlated as described below. At each time $t>t_0$, there are several $7 \times 7$ blocks having constant pixel value $5$. All other entries in $S_t$ are zero. Each block can either stay static with probability $0.8$ or move independently one pixel step towards top/bottom/left/right with probability $0.05$.
\end{itemize}

For ReProCS, we use the first $t_0=10^{4}$ frames, $[M_1, \cdots M_{t_0}] =[L_1, \cdots L_{t_0}]$, as training data to get an initial PCs' estimate $\hat{P}_{t_0}$ via computing its SVD. For $t > t_0$, we do Algorithm \ref{algo_CS} to recursively estimate $L_t$ and $S_t$. The support estimation threshold in (\ref{supp}), $\gamma$, can be set a little lower than the minimum nonzero magnitude that we would like to correctly detect. We set $\gamma =  0.2 \min_{i \in T_t} |(S_t)_i| =  1$. 
We update $\hat{P}_t$ for every $\tau = 20$ frames by Algorithm \ref{updatePt} with $\alpha=5$. The reconstruction time of ReProCS is about $0.015$ seconds per frame. To make a fair comparison, we use the entire sequence, $[M_1, \cdots M_{t_0+200}]$, to do PCP\cite{rpca} and RSL\cite{Torre03aframework}. The reconstruction time of offline PCP, $\text{(total time)}/(t_0+200)$, is about $0.05$ seconds per frame\footnote{PCP using the last $200$ frames takes about $0.02$ seconds per frame but it gives even larger errors for $L_t$ and $S_t$}. The reconstruction time of offline RSL, $\text{(total time)}/(t_0+200)$, is about $0.03$ seconds per frame. But if we had to do PCP or RSL for a causal reconstruction, i.e., at each time $t$, estimate $S_t$ and $L_t$ by solving PCP or RSL using all available data $[M_1, \cdots, M_t]$, PCP will take about $500$ seconds and RSL will take about $300$ seconds at each time.

\begin{table*}
\centerline{
\subtable[$\text{rank}(L) = 36$, $\text{supp}(S)$ is uniformly random] {
{\renewcommand{\arraystretch}{1.2}
    \begin{tabular}{ | l | c | c | c | c |}
    \hline
    \multirow{2}{*}{$\frac{|T_t|}{n}$} & \multicolumn{2}{|c|}{PCP} & \multicolumn{2}{|c|}{ ReProCS } \\
    \cline{2-5}
    & $ \frac{\mathbb{E}\|L - \hat{L}\|_F^2}{\mathbb{E}\|L\|_F^2}$ & $\frac{\mathbb{E}\|S - \hat{S}\|_F^2}{\mathbb{E}\|S\|_F^2}$
    & $ \frac{\mathbb{E}\|L - \hat{L}\|_F^2}{\mathbb{E}\|L\|_F^2}$ & $\frac{\mathbb{E}\|S - \hat{S}\|_F^2}{\mathbb{E}\|S\|_F^2}$ \bigstrut \\
    \hline
    $9.57\%$  & $1.42 \times 10^{-4}$  & $3.0 \times 10^{-3}$  & $1.32 \times 10^{-4}$  & $2.8 \times 10^{-3}$  \\
    \hline
    $19.14\%$ & $5.58 \times 10^{-4}$  & $5.8 \times 10^{-3}$  & $1.81 \times 10^{-4}$  & $1.9 \times 10^{-3}$  \\
    \hline
    $28.71\%$ & $1.80 \times 10^{-3}$  & $1.25 \times 10^{-2}$ &$2.42 \times 10^{-4}$ & $1.7 \times 10^{-3}$  \\
    \hline
    \end{tabular} \label{table_random_32}
}
}
\subtable[$\text{rank}(L) = 36$, $\text{supp}(S)$ is correlated]{
{\renewcommand{\arraystretch}{1.2}
    \begin{tabular}{ | l | c | c | c | c |}
    \hline
    \multirow{2}{*}{$\frac{|T_t|}{n}$} & \multicolumn{2}{|c|}{PCP} & \multicolumn{2}{|c|}{ ReProCS } \\
    \cline{2-5}
    & $\frac{\mathbb{E}\|L - \hat{L}\|_F^2}{\mathbb{E}\|L\|_F^2}$ & $\frac{\mathbb{E}\|S - \hat{S}\|_F^2}{\mathbb{E}\|S\|_F^2}$
    & $\frac{\mathbb{E}\|L - \hat{L}\|_F^2}{\mathbb{E}\|L\|_F^2}$ & $\frac{\mathbb{E}\|S - \hat{S}\|_F^2}{\mathbb{E}\|S\|_F^2}$ \bigstrut \\
    \hline
    $9.57\%$ & $9.6 \times 10^{-4}$ & $0.022$ & $2.80\times 10^{-4}$  & $6.4\times 10^{-3}$ \\
    \hline
    $19.14\%$ & $2.76 \times 10^{-2}$ & $0.333$ & $4.75 \times 10^{-4}$ & $5.7 \times 10^{-3}$  \\
    \hline
    $28.71\%$ & $5.81 \times 10{-2}$ & $0.502$ & $6.52\times 10^{-4}$ & $5.6\times 10^{-3}$  \\
    \hline
    \end{tabular} \label{table_structured_32}
}
}
}
\centerline{
\subtable[$\text{rank}(L) = 132$, $\text{supp}(S)$ is uniformly random] {
{\renewcommand{\arraystretch}{1.5}
    \begin{tabular}{ | l | c | c | c | c |}
    \hline
    \multirow{2}{*}{$\frac{|T_t|}{n}$} & \multicolumn{2}{|c|}{PCP} & \multicolumn{2}{|c|}{ ReProCS } \\
    \cline{2-5}
    & $ \frac{\mathbb{E}\|L - \hat{L}\|_F^2}{\mathbb{E}\|L\|_F^2}$ & $\frac{\mathbb{E}\|S - \hat{S}\|_F^2}{\mathbb{E}\|S\|_F^2}$
    & $ \frac{\mathbb{E}\|L - \hat{L}\|_F^2}{\mathbb{E}\|L\|_F^2}$ & $\frac{\mathbb{E}\|S - \hat{S}\|_F^2}{\mathbb{E}\|S\|_F^2}$ \bigstrut \\
    \hline
    $9.57\%$  & $1.69 \times 10^{-4}$  & $1.32 \times 10^{-2}$  & $4.98 \times 10^{-5}$  & $3.9 \times 10^{-3}$  \\
    \hline
    $19.14\%$ & $6.55 \times 10^{-4}$  & $2.55\times 10^{-2}$   & $6.89 \times 10^{-5}$  & $2.7 \times 10^{-3}$  \\
    \hline
    $28.71\%$ & $1.70\times 10^{-3}$   & $4.33 \times 10^{-2}$  &$1.05 \times 10^{-4}$ & $2.7 \times 10^{-3}$  \\
    \hline
    \end{tabular} \label{table_random_128}
}
}
\subtable[$\text{rank}(L) = 132$, $\text{supp}(S)$ is correlated]{
{\renewcommand{\arraystretch}{1.5}
    \begin{tabular}{ | l | c | c | c | c |}
    \hline
    \multirow{2}{*}{$\frac{|T_t|}{n}$} & \multicolumn{2}{|c|}{PCP} & \multicolumn{2}{|c|}{ ReProCS } \\
    \cline{2-5}
    & $\frac{\mathbb{E}\|L - \hat{L}\|_F^2}{\mathbb{E}\|L\|_F^2}$ & $\frac{\mathbb{E}\|S - \hat{S}\|_F^2}{\mathbb{E}\|S\|_F^2}$
    & $\frac{\mathbb{E}\|L - \hat{L}\|_F^2}{\mathbb{E}\|L\|_F^2}$ & $\frac{\mathbb{E}\|S - \hat{S}\|_F^2}{\mathbb{E}\|S\|_F^2}$ \bigstrut \\
    \hline
    $9.57\%$ & $4.61 \times 10^{-4}$ & $0.04 $ & $1.03 \times 10^{-4}$  & $9.1 \times 10^{-3}$ \\
    \hline
    $19.14\%$ & $7.5 \times 10^{-3}$ & $0.35$ & $1.74 \times 10^{-4}$ & $8.1 \times 10^{-3}$  \\
    \hline
    $28.71\%$ & $1.64 \times 10^{-2}$ & $0.53$ & $2.47 \times 10^{-4}$ & $8.1 \times 10^{-3}$  \\
    \hline
    \end{tabular} \label{table_structured_128}
}
}
}
\vspace{-0.15in}
\caption{\small{Comparison of reconstruction errors of PCP\cite{rpca} and ReProCS for separating low rank $L$ and sparse $S$. Here, $L$ and $S$ are of size $n \times 200$ and $n = 1024$; the rank of $L$, $\text{rank}(L)$, is much smaller than $n$; and the support of $S$ is generated either uniformly at random or correlated. Note that $|T_t|$ is the number of nonzero elements in $S_t$.
}}\label{table1}
\vspace{-0.15in}
\end{table*}

\begin{figure*}
\centerline{
\subfigure[Normalized MSE of $L_t$ from $t_0+1$ to $t_0+200$]{
\psfig{file = 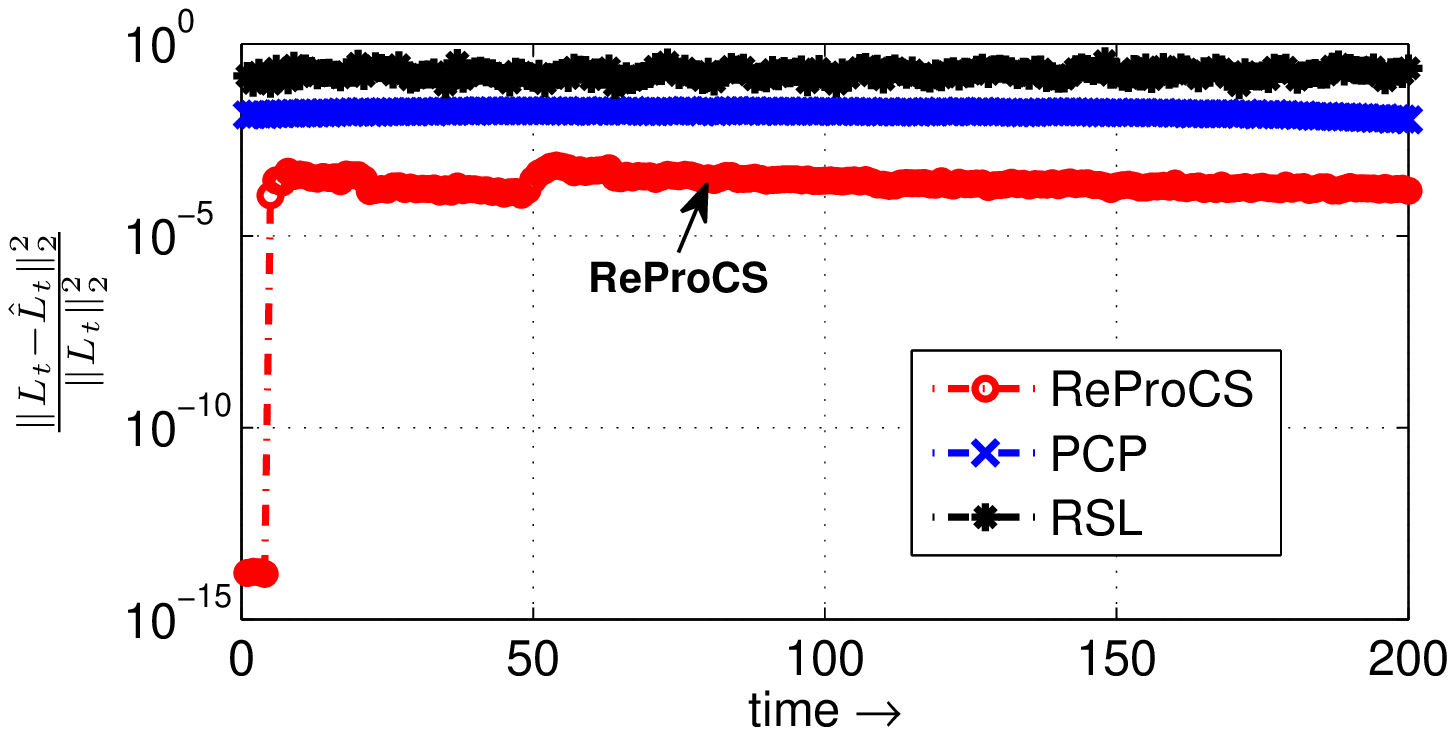, width=8cm} \label{expt1_errL}
}
\subfigure[Normalized MSE of $S_t$ from $t_0+1$ to $t_0+200$]{
\psfig{file = 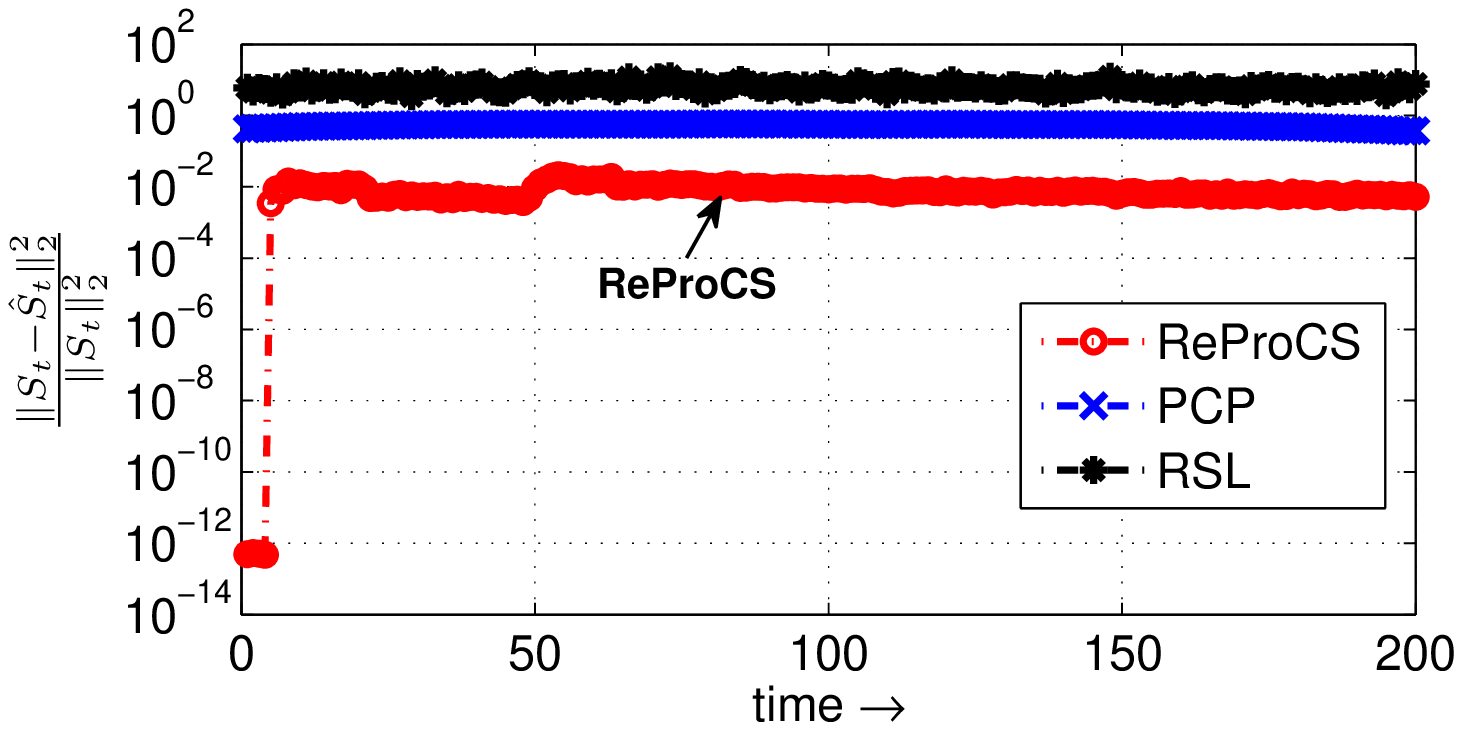, width=8cm} \label{expt1_errS}
}
}
\centerline{
\subfigure[Recovery results at $t=t_0+200$. The original observed image is $M_t = L_t + S_t$.]{
\psfig{file = 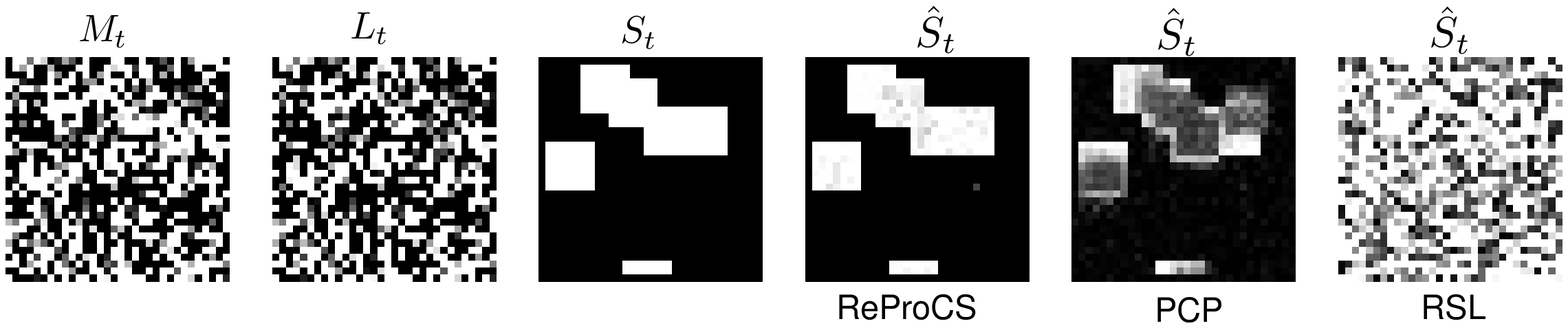, width=12cm} \label{expt1_image}
}
}
\vspace{-0.15in}
\caption{\small{Normalized MSE plots and reconstruction results for the third case in Table \ref{table1}\ref{table_structured_128} where $\frac{|T_t|}{n} \approx 28.71\%$}}\label{expt1}
\vspace{-0.15in}
\end{figure*}

We summarize the reconstruction errors of PCP and ReProCS in Table \ref{table1} based on $100$ times Monte Carlo averaging. The normalized MSE $\mathbb{E}(\|L-\hat{L}\|_F^2) / \mathbb{E}(\|S\|_F^2)$ and $\mathbb{E}(\|S-\hat{S}\|_F^2) / \mathbb{E}(\|S\|_F^2)$ are computed only for the last $200$ frames.
As can be seen from Table \ref{table1}, PCP has large reconstruction error for less sparse and/or correlated $S_t$'s. However, ReProCS can recursively recover $L_t$ and $S_t$ accurately, with reconstruction error less than $10^{-2}$ in all cases.
In Fig. \ref{expt1_errL} and Fig. \ref{expt1_errS}, we plot the normalized MSE (NMSE) of $L_t$ and $S_t$ for the third case in Table \ref{table1}\ref{table_structured_128} where $\frac{|T_t|}{n} \approx 28.71 \%$. We also show the error of RSL. Since the nonzero elements of $S_t$ have much smaller magnitude than the elements of $L_t$, RSL also does not work. In Fig. \ref{expt1_image}, we show $\hat{S}_{t_0+200}$ obtained by ReProCS, PCP,  RSL \cite{Torre03aframework} as an image. 

We also verified that ReProCS keeps tracking the change of PCs gradually and correctly. At $t=t_0$, we have a good $\hat{P}_{t_0}$ that finds all the existing PCs. At $t=t_0+5$, let $U_1$ denote the two new PCs that get added and let $U_2$ denote the two old PCs that start to be removed from $\text{span}(\Phat_t)$. At $t=t_0$, we have $\frac{\|\hat{P}_{t}' U_1\|_F^2}{\|U_1\|_F^2} =0$ ($U_1$ is not in $\text{span}(\hat{P}_t)$) and $\frac{\|\hat{P}_{t}' U_2\|_F^2}{\|U_2\|_F^2} = 1$ ($U_2$ is in $\text{span}(\hat{P}_t)$). We updated $\hat{P}_t$ for every $\tau = 20$ frames using Algorithm \ref{updatePt} with $\alpha = 0.5\sigma_{\text{min}}^2 = 5$. 
At $t=t_0+200$, $\frac{\|\hat{P}_{t}' U_1\|_F^2}{\|U_1\|_F^2}$ increased from $0$ to $0.91$ and $\frac{\|\hat{P}_{t}'U_2\|_F^2}{\|U_2\|_F^2}$ decreased from $1$ to $0.11$. Therefore, new PCs gradually and correctly got added to $\text{span}(\hat{P}_t)$ and the decayed PCs gradually got removed from $\text{span}(\hat{P}_t)$. As explained in Sec. \ref{sec_updatePt}, we can remove the decayed PCs more quickly by using a larger threshold $\alpha$.

%

\subsection{Using correlation of the sparse part: the need for supp-pred-modCS and  ReProCS (modCS)}\label{expt_modcs}

In this section, we show an example where ReProCS (modCS) is needed.  The image sequence is of size $64 \times 80 \times (t_0 + 100)$ with $t_0 =5000$. 
Thus, the measurement at time $t$, $M_t$, is of length $n = 64 \times 80 = 5120$. We generate $M_t = L_t + S_t$ as follows.  We set $L_t = U x_t$ where $U$ is generated as before and $x_t$ follows the regression model of the Appendix with $f=0.5$, $f_d = 0.1$, and $\theta=0.5$. Initially, there are $0.2n$ PCs with variance $10^4, 0.9933 \times 10^4, 0.9933^3\times 10^4, \cdots, 10$. At $t=t_0+5$, two new PCs with variance $50$ and $55$ get added and the value of $x_t$ along two old PCs starts to decay to zero. Compared to the previous experiment, the rank of the PC matrix, $r = \text{rank}(\hat{P}_t) \approx 0.2 n$, is larger.
For $1 \le t \le t_0$, $S_t=0$ and hence $M_t=L_t$. For $t>t_0$, there are two $45 \times 29$ moving objects in $S_t$. Thus, the support size of $S_t$ is $|T_t| \approx 0.51 n$ while the number of projected measurements, $y_t$, is $n-r \approx 0.8 n$. The two blocks move independently following the motion model (\ref{motion_model}) where $n_t$ is a zero mean truncated Gaussian noise with variance $Q = 2.5 \times 10^{-5}$, i.e., $n_t \sim \mathcal{N} (0,Q)$ and $-2\sqrt{Q}<|n_t|<2\sqrt{Q}$. The first block has constant pixel intensity $10$ and it moves from left to right with initial velocity $v_{t_0+1} = 0.25$. The second block has constant intensity $20$ and it moves from right to left with $v_{t_0+1} = -0.25$.
We use a small $Q$ because we want the blocks to stay in the scene for a long sequence (in the next section, we use a larger $Q$).%

For ReProCS(modCS), at time $t=t_0 + 1$, we start a separate KF for tracking the motion of each object with its true location and velocity, i.e. we use $\hat{g}_{t_0+1|t_0} = [p_{t_0+1}, \ v_{t_0+1}]$, $\Sigma_{t_0+1|t_0} = 0$ for each object.
The deletion threshold, $\alpha_{\text{del}}$, can be set a little lower than the minimum nonzero magnitude that we would like to correctly detect. Thus we set $\alpha_{\text{del}} = a \min_{i \in N_t} |(S_t)_i| = 10a$ for an $a<1$. We used $a=0.1$ and thus set $\alpha_{\text{del}}=1$. The addition threshold usually can be much lower to ensure most elements get correctly added. This  can either be indirectly set (we can keep adding more elements to $T_{\text{add}}$ until the condition number of $(A_t)_{T_{\text{add}}}$ goes below a ``well-conditioned" threshold) or we can just set it to a percentage of $\alpha_{\text{del}}$. In this work we used this latter approach and set $\alpha_{\text{add}}=0.5 \alpha_{\text{del}}$. To obtain the observed locations of the different objects, we first use intensity thresholds to obtain their respective supports. The observed location of each block is then obtained as the median of its estimated support, which is more robust than (\ref{supportmean}) if there are occasionally some extra indices far away from the true support set. We use $R=4Q=10^{-4}$ as the variance of the observation noise, $w_t$, in (\ref{obsmod4}). We use a larger $R$ because the intensities of the moving objects are small compared to the background and hence the support update error is likely to be larger.
For both ReProCS and ReProCS(modCS), we update $\hat{P}_t$ for every $\tau = 20$ frames with $\alpha = 0.5 \sigma_{\text{min}}^2=5$. For ReProCS, we use $\gamma=1$ for support thresholding in (\ref{supp}).%

\begin{figure*}
\centerline{
\subfigure[Reconstruction Error of $S_t$ from $t_0+1$ to $t_0+100$]{
\psfig{file = 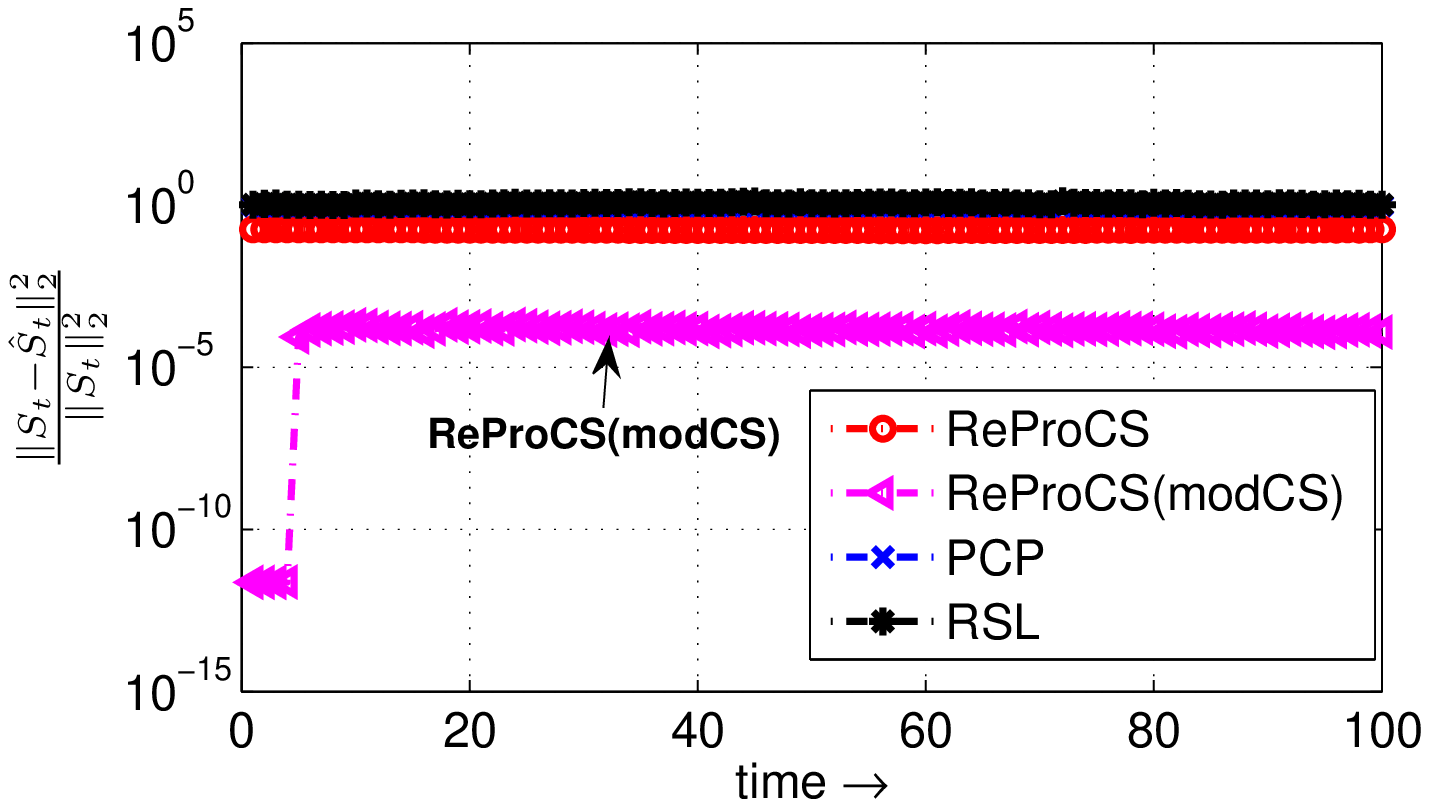, height=4.5cm, width=7.5cm} \label{expt3_errS}
}
\subfigure[Support error of supp-pred-modCS]{
\psfig{file = 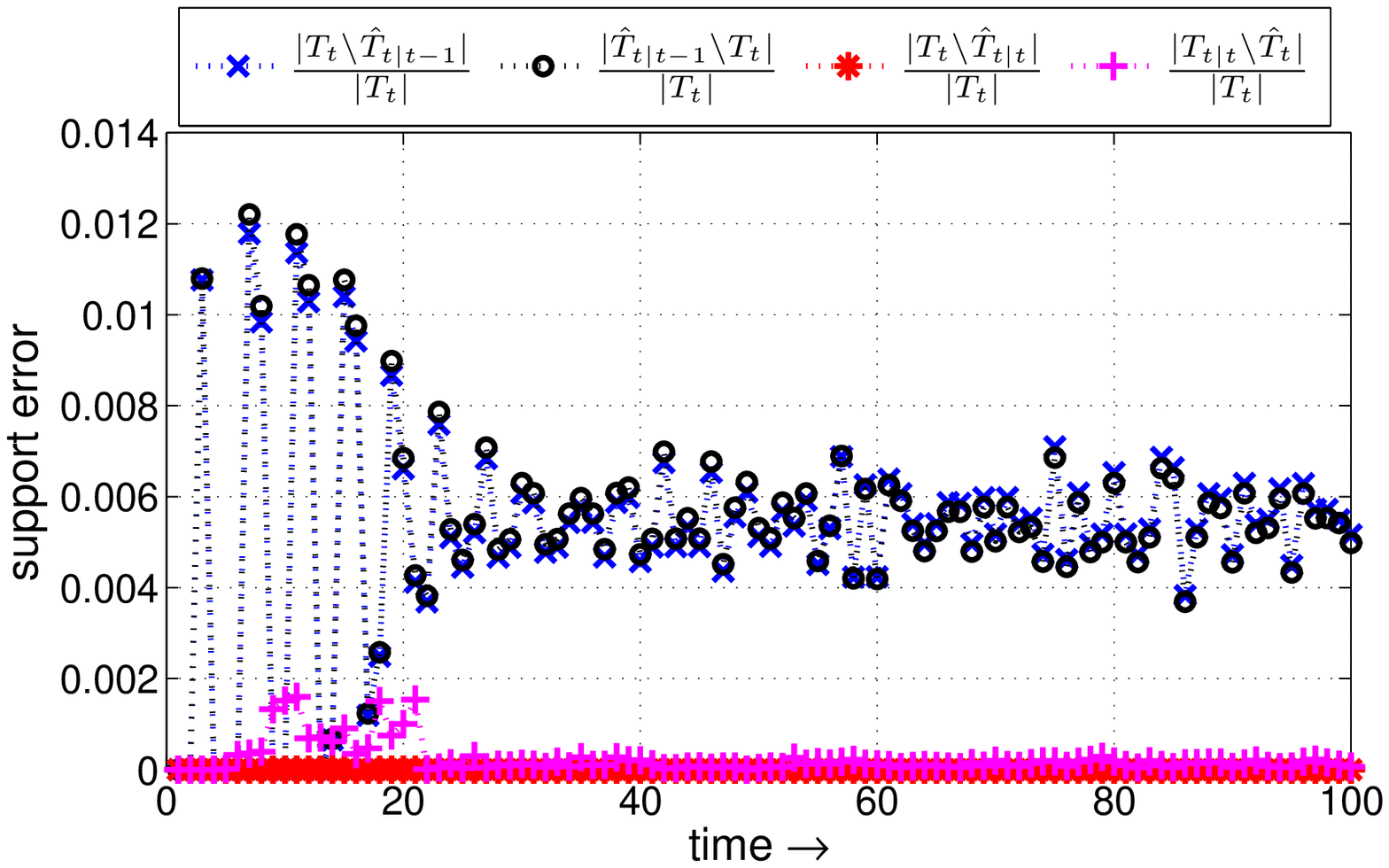, height=4.5cm, width=7.5cm} \label{expt3_supporterror}
}
}
\centerline{
\subfigure[Recovered image $S_t$ at $t=t_0+100$]{
\psfig{file = 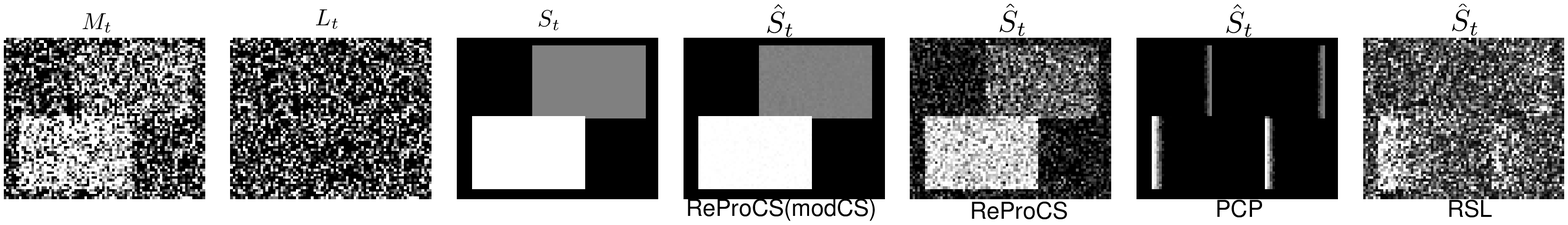, width=15cm} \label{expt3_image}
}
}
\vspace{-0.15in}
\caption{\small{Comparison of ReProCS and ReProCS(modCS) when limited projected measurements are available. Here, $|T_t| = \ |\text{supp}(S_t)| \approx 0.51 n$ while the number of projected measures is about $n - \text{rank}(P_t) \approx 0.8n$.}}\label{expt3_two}
\vspace{-0.15in}
\end{figure*}

In practice, foreground objects slowly enter the scene and the support size of $S_t$, $|T_t|$, increases over time (until they enter the scene completely). Therefore, $|T_t|$ shall be small for the initial frames. Even though we may not have the initial location/velocity knowledge, we can do ReProCS for the first a few frames, get initial estimates of the support and then use intensity based segmentation followed by centroid computation to estimate the initial location/velocity of the objects. For the future frames when $|T_t|$ is large, we can replace CS by supp-pred-modCS, i.e.,  use ReProCS(modCS).%

In Fig. \ref{expt3_errS}, we plot the NMSE of $S_t$ for ReProCS and ReProCS(modCS) based on $100$ times Monte Carlo averaging. We also plot the reconstruction error of PCP and RSL for one realization. ReProCS takes about $2.1$ seconds per frame and ReProCS(modCS) takes about $2.8$ seconds per frame.
As can be seen, with a large $\frac{|T_t|}{n-r}$, ReProCS(modCS) outperforms ReProCS greatly because it utilizes the correlation model of $S_t$ while ReProCS does not. PCP and RSL again do not work. 
%
In Fig. \ref{expt3_supporterror}, we plot the average number of extras and misses in $\hat{T}_{t|t-1}$ (predicted support) and $\hat{T}_{t|t}$ (updated support) used by supp-pred-modCS. 
Clearly, modCS corrects most prediction errors.%

\subsection{ReProCS and ReProCS(modCS) on a real background sequence with simulated foreground images} \label{expt_wave}
\begin{figure*}
\centerline{
\subfigure[Reconstruction Error of $S_t$]{
\psfig{file = 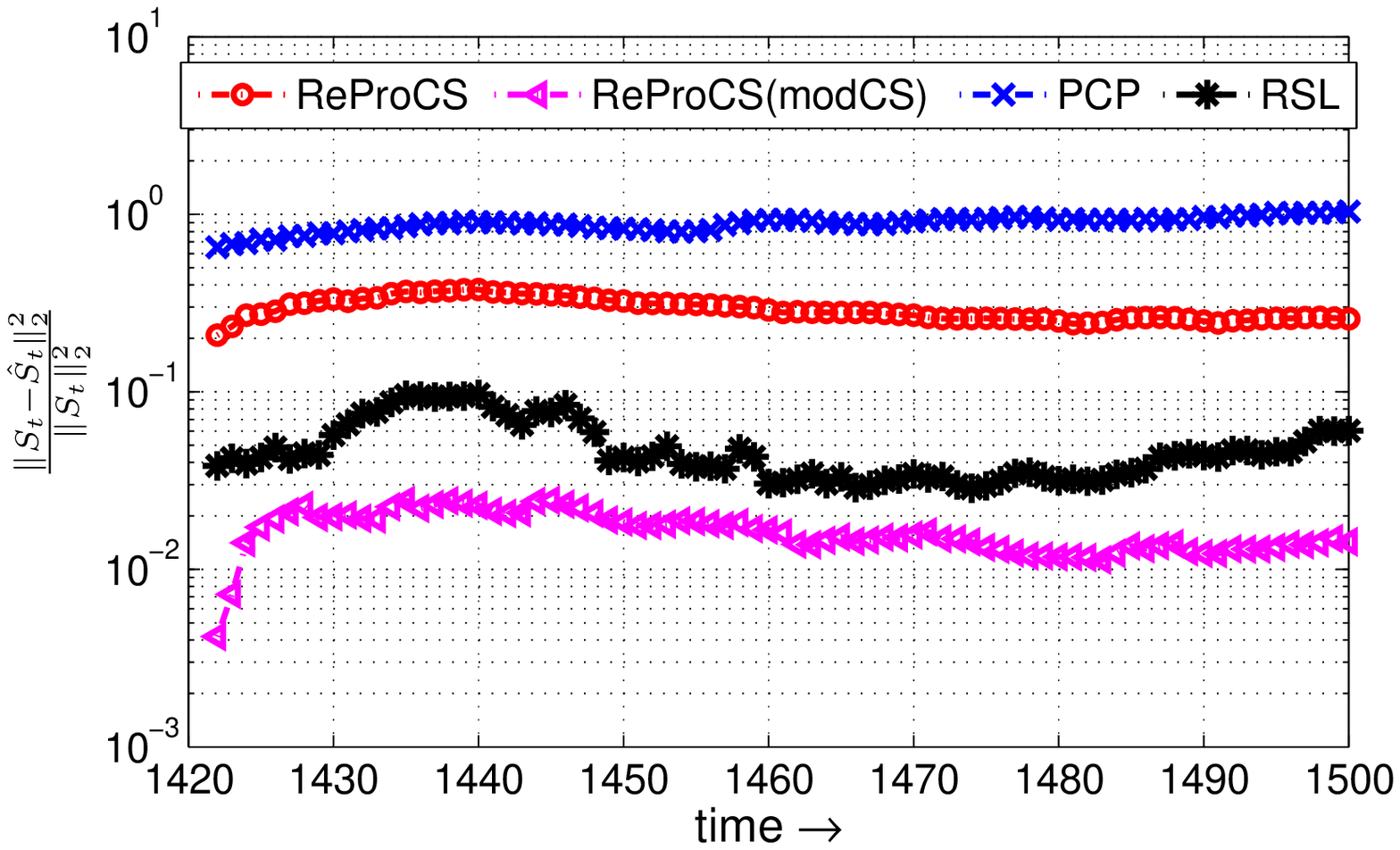, height = 4 cm,width=7.5cm} \label{expt4_errS}
}
\hspace{-0.1in}
\subfigure[Reconstruction Error of $O_t$]{
\psfig{file = 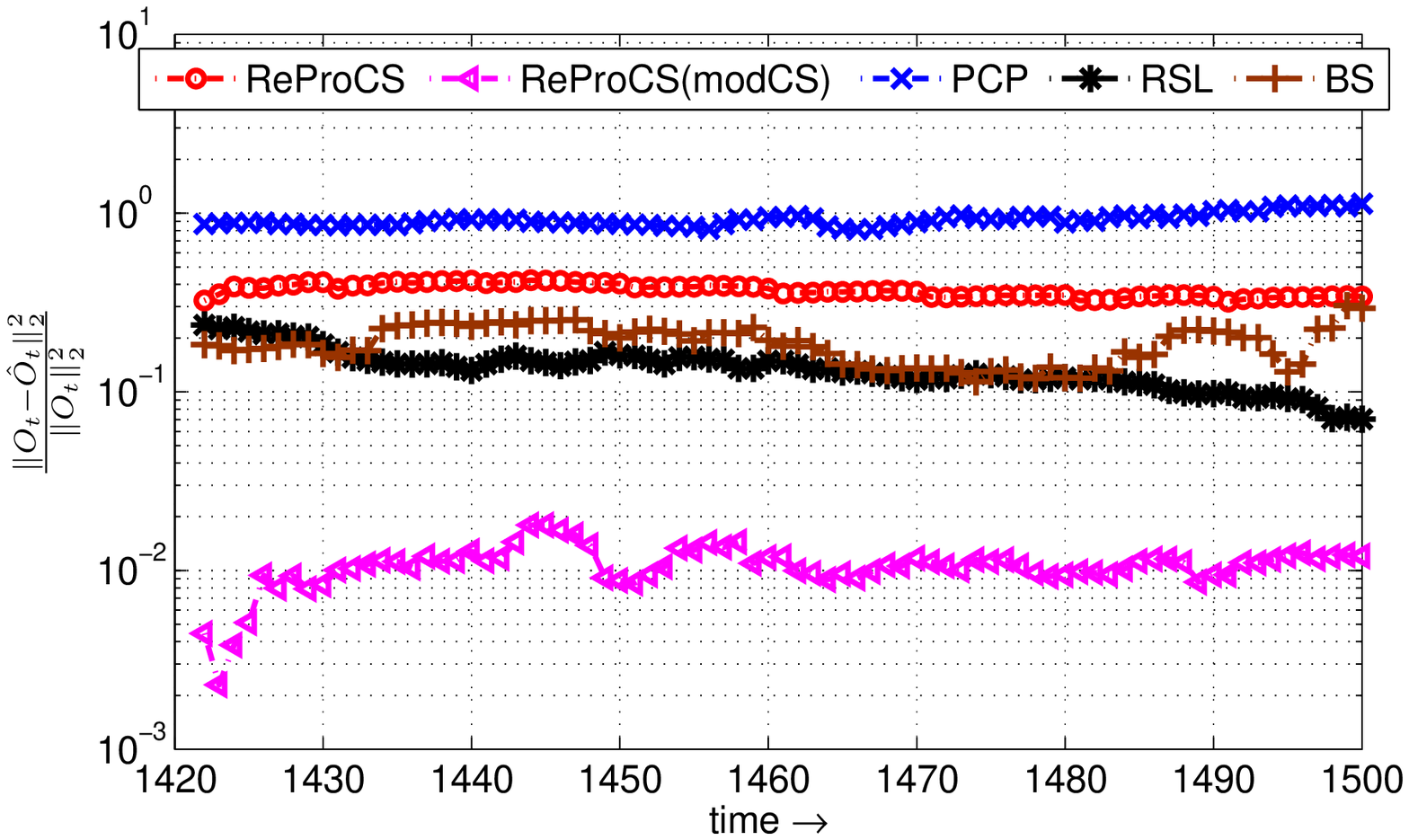, height = 4 cm,width=7.5cm} \label{expt4_errO}
}
}
\centerline{
\subfigure[First row: original image at $t=t_0+20$. Second row: recovered foreground image ]{
\psfig{file = 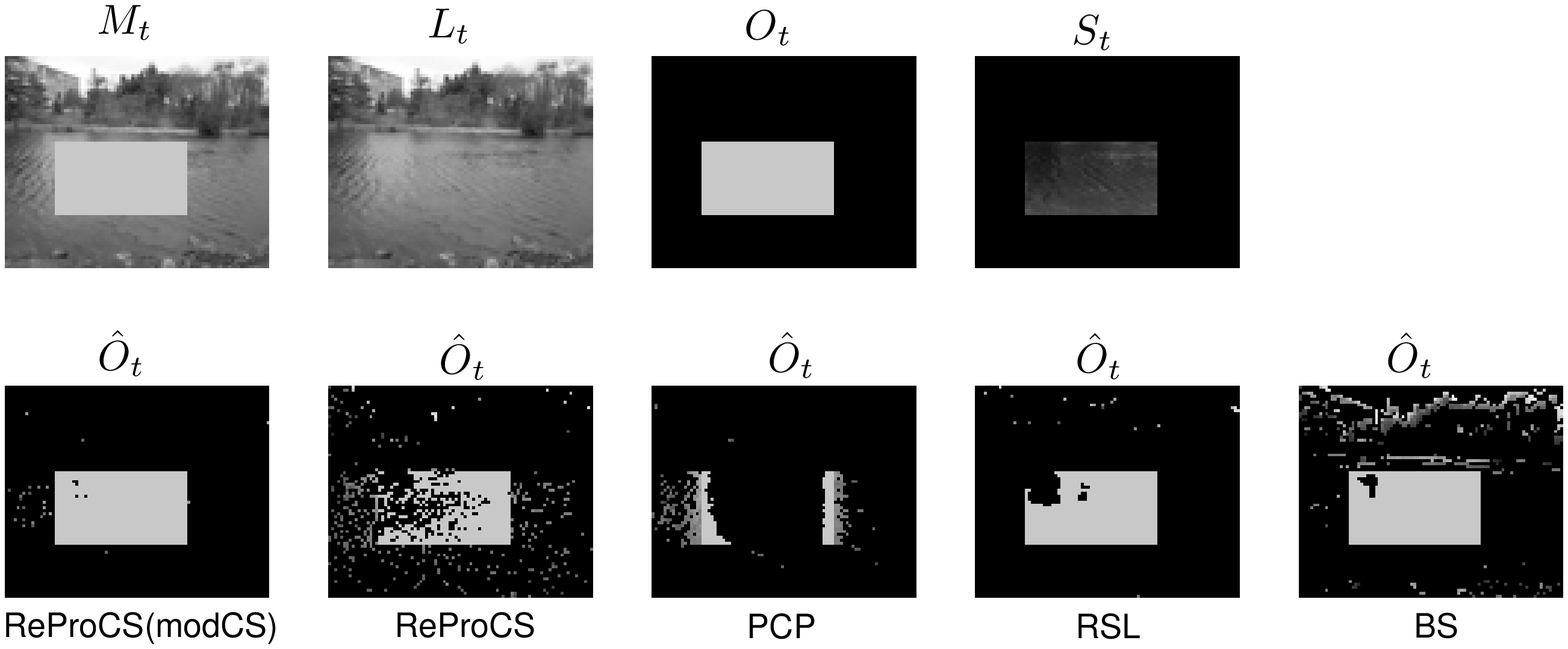,width=12cm} \label{expt_wave_org}
}
}
\vspace{-0.15in}
\caption{\small{Here, $\text{rank}(P_t) \approx 0.2n$, and $|T_t| =  \ |\text{supp}(S_t)| \approx 0.16n$.}}\label{expt4_large}
\vspace{-0.15in}
\end{figure*}

In this experiment, we compare ReProCS and ReProCS(modCS) with PCP, RSL, and thresholding based background subtraction (BS) on a partly-real video sequence. We take a $72 \times 90 \times 1500$ real video sequence around a lake with slow global background variations, e.g., water waves.
By arranging each image frame as a $n = 72 \times 90 = 6480$ dimensional column vector, we get the background sequence, $[L_1,\cdots,L_{1500}]$, which has nonzero mean. The first $t_0=1420$ frames serve as the training data. The nonzero background mean, denoted by $\mu_0$, is estimated as the empirical mean of the training sequence, i.e., $\mu_0 = (L_1+\cdots+ L_{t_0})/t_0$.  
We subtract $\mu_0$ from $L_t$ and then get the initial PCs estimate, $\hat{P}_{t_0}$, by computing a SVD decomposition of $[L_1 - \mu_0,\cdots L_{t_0}-\mu_0]$. We get $r = \text{rank}(\hat{P}_{t_0}) = 1255 \approx 0.2n$. For $1 \le t \le t_0$, we let $M_t = L_t$. For $t>t_0$, we {\em overlay} a simulated foreground image, $O_t$, on the background image $L_t$. The observed image, $M_t$, is thus $(M_t)_{T_t} = (O_t)_{T_t}$ and $(M_t)_{T_t^c} = (L_t)_{T_t^c}$. Thus, $S_t$ follows (\ref{S_def}). 
The foreground image has one $45 \times 25$ moving block with nonzero pixel intensity $200$. The foreground block moves slowly from left to right following the motion model (\ref{motion_model}) with initial velocity $v_{t_0+1} = 0.5$ and velocity change variance $Q = 0.005$ ($n_t$ is truncated Gaussian noise with $n_t \sim \mathcal{N} (0,Q)$ and $-2\sqrt{Q}<|n_t|<2\sqrt{Q}$). We show $M_t$, $L_t$, $O_t$, and $S_t$ (shown in 2D fashion) at $t=t_0+20$ in Fig. \ref{expt_wave_org}. Compared with the experiments in Sec. \ref{expt_comparison} and Sec. \ref{expt_modcs}, the magnitudes of most of the nonzero elements of $S_t$'s are now larger but they vary over location and time. In this case $\min_{i \in T_t} |(S_t)_i| \approx 25$.

Note that $L_t$'s have nonzero mean $\mu_0$ and $\hat{P}_t$ estimates the PCs of $[L_1-\mu_0, \cdots, L_t-\mu_0]$ instead of $[L_1, \cdots, L_t]$. We do ReProCS and ReProCS(modCS) on $M_t-\mu_0$ to estimate $\hat{S}_t$ and then let $\hat{L}_t = M_t - \hat{S}_t$. We do Algorithm \ref{updatePt} to update $\hat{P}_t$ for every $10$ frames with $D \leftarrow [D \  \hat{L}_t-\mu_0]$ and $\alpha= 0.1$.
For ReProCS, we use $\gamma=10$ for support thresholding in (\ref{supp}).  For ReProCS(modCS), we run a KF in the same way as in Sec. \ref{expt_modcs}. We use $R=Q/50 = 10^{-4}$ in (\ref{obsmod4}) since $S_t$'s now have larger magnitude and so we expect the modCS output to be more reliable. We set $\alpha_{\text{del}}=0.8 \min_{i \in T_t} |(S_t)_i| = 20$ and $\alpha_{\text{add}}=0.5 \alpha_{\text{del}}=10$. 
We do PCP and RSL using the entire sequence, $[M_1, \cdots,  M_{1500}]$, and get $\hat{S}_t$. The support estimation, $\hat{T}_t$, is estimated as the $90\%$-percent energy set of $\hat{S}_t$. The foreground estimation is $(\hat{O}_t)_{\hat{T}_t} = (M_t)_{\hat{T}_t}$, $(\hat{O}_t)_{\hat{T}_t^c}=0$.
%
%
BS first subtracts the background mean, $\mu_0$, from the current image observation, $M_t$, and then estimate $O_t$ by thresholding on the difference image, $M_t-\mu_0$, using a threshold picked in the same way for PCP or RSL.

As can been seen from Fig. \ref{expt4_large}, the reconstruction errors of PCP, RSL, BS and ReProCS are all much larger than ReProCS(modCS). The recovered image frames at $t=20$ in shown in Fig. \ref{expt_wave_org}. 
In this experiment, the support size of $S_t$ is about $|T_t| \approx 0.16n$ and the number of projected measurements is about $n-r \approx 0.8n$, which seems to be enough for recovering $S_t$ using ReProCS. However, as shown in the figure, ReProCS cannot recover $S_t$ very well. This is because that there is a lot of background variation hence the projection matrix, $A_t = (\hat{P}_{t,\perp})'$, cannot make the noise $\beta_t$ sufficiently small compared to $S_t$. When the noise $\beta_t$ is large, for a given support size $|T_t|$, more measurements are needed for accurate sparse recovery.
We did the same experiment but with a smaller foreground block of size $25 \times 19$ ($|T_t| \approx 0.07n$). In that case, ReProCS and ReProCS(modCS) give similar reconstruction errors which are much smaller than PCP, RSL, and BS (not shown).

In Fig. \ref{expt_wave_org}, there are some misses and extras in $\hat{O}_t$ estimated by ReProCS(modCS). This is because the magnitudes of nonzero elements of $S_t$ vary over location and time. 
Also, as explained earlier, the noise $\beta_t$ may not be sufficiently small as compared with the small elements of $S_t$. 
A possible solution is to use location and time varying thresholds. Also, using the fact that $S_t$ is block sparse can significantly help improve performance.

\section{Conclusions, Limitations and Future Work}\label{sec_conclusion}
This work finds the missing link between the recursive robust PCA problem and the recursive sparse recovery problem in large noise. From the robust PCA perspective, our proposed solutions, ReProCS and ReProCS(modCS), are robust to correlated ``outliers" as long as they are sparse. From the sparse recovery perspective, they are robust to large ``noise" as long as the ``noise" is spatially correlated enough to have an approximately low rank covariance matrix and this covariance matrix is either constant or changes slowly with time. Our second solution, ReProCS(modCS), which utilizes the correlation model of the sparse part, can handle significantly less sparse $S_t$'s or significantly larger ranks of the low rank part.   

A limitation of our approach, and of recursive robust PCA in general, is that it requires an initial estimate of the PC matrix, $\hat{P}_0$. To get that we either need to know that a certain initial set of frames have no sparse part, or we need to use PCP to recover the low rank part and hence estimate $\hat{P}_0$. Additionally, it also requires that the $L_t$'s are spread out enough (not sparse) and that the low dimensional subspace in which  $L_t$ lies changes slowly over time. ReProCS(modCS) also needs to know the structure of the correlation model on the sparse part.
%

In this work we study the case of recursively recovering a sparse vector from full measurements, i.e. $M_t = S_t + L_t$, when the corrupting noise has large magnitude, but is highly spatially correlated (has a low rank covariance matrix). But both ReProCS and ReProCS(modCS) will extend directly to the more general case of  $M_t = \Psi S_t + L_t$ where $\Psi$ may be a fat or a square matrix. The only change will be that we will need to use $A_t = (\hat{P}_{t,\perp})' \Psi$ instead of just $A_t = (\hat{P}_{t,\perp})'$ in both Algorithm \ref{algo_CS} and Algorithm \ref{algo_modCS}. Also, we will need the assumption that $(\hat{P}_{t,\perp})' \Psi$ does not nullify the sparse vectors $S_t$.

This work introduces the idea of {\em support-predicted modified-CS} using one very simple correlation model. Similar ideas can be extended to many more general models and to many applications other than video surveillance. 
%
Also, in certain cases, the support change may be slow enough so that the previous support estimate may itself be a very good prediction of the current support.  In this case the correlation model update using a KF is not needed. For the video application, certain practical issues are discussed in Sec. \ref{extension_modCS}. Moreover, so far we only use support prediction. We can also try to use (a) signal value prediction and (b) incorporate spatial correlation, e.g. using \cite{modelcs}. Finally, the arguments to obtain the conditions under which stability (time-invariant and small bound on the error) of support-predicted-modified-CS holds need to be formalized in future work.

\section*{Appendix: Model on the support change of $x_t$}

We provide here a realistic generative model for $x_t$ and hence the low rank part, $L_t = Ux_t$ that satisfies the assumptions given in Sec. \ref{intro_prob}. 
The support set of $x_t$, $N_t$, is a union of three disjoint sets $\mathcal{A}_t$, $\mathcal{D}_t$, and $\mathcal{E}_t$, i.e., $N_t = \mathcal{A}_t \cup \mathcal{D}_t \cup \mathcal{E}_t$.
At $t=1$, we let $N_t = \mathcal{E}_t$ and $\mathcal{A}_t \cup \mathcal{D}_t = \emptyset$.
For $t>1$, the new addition set $\mathcal{A}_t := N_t \setminus N_{t-1}$ is the set of indices along which $(x_t)_{\mathcal{A}_t }$ starts to be nonzero. $\mathcal{A}_t$ is nonempty once every $d$ frames.
The set $\mathcal{D}_t \subset (N_t \cap N_{t-1})$ is the set of indices along which $(x_t)_{\mathcal{D}_t }$ decay exponentially to zero. We assume that $\mathcal{D}_t$ will not get added to $N_t$ at any future time.
The set $\mathcal{E}_t := N_t \cap N_{t-1} \setminus \mathcal{D}_t$ is the set of indices along with $(x_t)_{\mathcal{E}_t }$ follows a first order autoregressive (AR-1) model.

Let $\Sigma$ be a diagonal matrix with nonnegative diagonal elements $\sigma_i^2$s. We model $x_t$ as
\begin{eqnarray}
&&x_0 = 0  \nonumber \\
&& x_t = F_t x_{t-1} + \nu_t, \ \nu_t \overset{\text{i.i.d.}}{\sim} \mathcal{N}(0,Q_t) \label{signal_model}
\end{eqnarray}
where $\nu_t$ is independent and identically distributed Gaussian noise with zero mean and diagonal covariance matrix $Q_t$; $F_t$ and $Q_t$ are two diagonal matrices defined as below

\begin{eqnarray*}
F_t  &=& \left[
  \begin{array}{cccc}
    0_{\mathcal{A}_t} & 0 & 0 & 0\\
    0 & (f I)_{\mathcal{E}_t} & 0 & 0 \\
    0 & 0 & (f_d I)_{\mathcal{D}_t} & 0\\
    0 & 0 & 0 & (0)_{N_t^c} \\
  \end{array}
\right] \\
Q_t  &=& \left[
  \begin{array}{cccc}
    \theta (\Sigma)_{\mathcal{A}_t} & 0 & 0 & 0\\
    0 & (1-f^2) (\Sigma)_{\mathcal{E}_t} & 0 & 0\\
    0 & 0 & (0)_{\mathcal{D}_t} & 0\\
    0 & 0 & 0& (0)_{N_t^c} \\
  \end{array}
\right]
\end{eqnarray*}
The three scalars $f$, $f_d$, and $\theta$ satisfy $0< f_d <f <1$ and $0 < \theta <1$.

From the model on $x_t$ given in (\ref{signal_model}), we notice the following:
\begin{itemize}

\item[a)] At time $t = j d $, $(x_t)_{\mathcal{A}_t}$ starts with
\begin{equation*}
(x_t)_{\mathcal{A}_t} \sim \mathcal{N} (0, \theta (\Sigma)_{\mathcal{A}_t}).
\end{equation*}
Small $\theta$ ensures that new directions get added at a small value and increase slowly.
$(x_t)_{\mathcal{D}_t}$ decays as \begin{equation*}
(x_t)_{\mathcal{D}_t} = f_d (x_t)_{\mathcal{D}_t}
\end{equation*}
$(x_t)_{\mathcal{E}_t}$ follows an AR-1 model with parameter $f$:
\begin{equation*}
(x_t)_{\mathcal{E}_t} = f (x_{t-1})_{\mathcal{E}_t} + (v_t)_{\mathcal{E}_t}
\end{equation*}

\item[b)] At time $t>jd$, the variance of $(x_t)_{\mathcal{A}_{jd}}$ gradually increases as
\begin{equation*}
(x_t)_i \sim \mathcal{N}( 0, (1 -(1-\theta)f^{2(t-jd)})  \Sigma_{i,i}), \ \ i \in \mathcal{A}_{jd}
\end{equation*}
Eventually, the variance of $(x_t)_{\mathcal{A}_{jd}}$ converges to $(\Sigma)_{\mathcal{A}_{jd}}$. We assume this becomes approximately true much before $t=(j+1)d$ (the next support change time).

\item[c)] At time $t>jd$, the variance of $(x_t)_{\mathcal{D}_{jd}}$ decays exponentially to zero as
\begin{equation*}
(x_t)_{\mathcal{D}_{jd}} \sim \mathcal{N} (0, f_d^{2(t-jd)} (\Sigma)_{\mathcal{D}_{jd}})
\end{equation*}
We assume that this has approximately decayed to zero much before $t=(j+1)d$.
\end{itemize}

For every $d$ frames, at $t=jd$, the values of $x_t$ along the new indices, $\mathcal{A}_t$, become nonzero and the values of $x_t$ along existing indices, $\mathcal{D}_t$, start to decay exponentially. The values of $x_t$ along all other existing indices follow an independent AR-1 model. After a short period $\Delta_d$ which is much less than $d$, the variances of $x_t$ along the new indices increase to some stable values and the values of $x_t$ along the decaying indices decayed to zero. Therefore, $N_t$ is piecewise constant from $t=jd + \Delta_d$ to $t=jd+d$.

\bibliographystyle{ieeepes}
\bibliography{tipnewpfmt}

\clearpage  
\section*{Supplementary Material}
The overall ReProCS algorithm is a recursive approach. In ReProCS(modCS) the sparse recovery algorithm, supp-pred-modCS, is itself recursive. Hence an important question is when and why will it be stable (error bounded by a time-invariant and small value)? Our simulations given in Sec. \ref{results} do indicate that it is stable. In this section, we provide the main arguments to show this analytically. These will be formalized in later work.
{\em Everywhere in the discussion below, ``bounded" means bounded by a time-invariant value.}

First consider supp-pred-modCS. We give here the main idea of an induction argument to show its stability. Suppose that at $t-1$, $\|S_{t-1} - \hat{S}_{t-1}\|_2$ is bounded and small and that the location error, $|p_{t-1} - \hat{p}_{t-1|t-1}|$ is also bounded and small. Since $n_t$ is assumed to be bounded and small, the second induction assumption means that $|p_{t} - \hat{p}_{t|t-1}|$ is also bounded and small. It is easy to see then that the same holds for the support prediction errors $|\Delta_{t|t-1}|:= |T_t \setminus \hat{T}_{t|t-1}|$ and $|\Delta_{e,t|t-1}|:= |\hat{T}_{t|t-1} \setminus T_t|$. In fact, both $|\Delta_{t|t-1}|$ and $|\Delta_{e,t|t-1}|$ are bounded by $|p_{t} - \hat{p}_{t|t-1}|$.
Using this and arguments similar to those in \cite{stability_allerton}, we should be able to argue that the sparse recovery and support update step (modified-CS with Add-LS-Del) will also result in (i) $\hat{T}_{t|t}$ with bounded and small number of extras, $|\Delta_{e,t|t}|$, and misses, $|\Delta_{t|t}|$ and (ii) the error of the recovered sparse part, $\|S_t - \hat{S}_t\|_2$ being bounded and small. This step will need to assume that (a) the noise seen by modified-CS, $\beta_t$, is bounded and small; (b) $A_t$ satisfies a certain RIP condition (if $|\Delta_{t|t-1}|\le a$ and $|\Delta_{e,t|t-1}|\le b$ then we will need $\delta_{2w+1+a+b}< (\sqrt{2}-1)$) \cite{candes_rip,stability_allerton}); and (c) most nonzero elements of $S_t$ are large enough. In fact using this approach, the bound on the final number of extras, $|\Delta_{e,t|t}|$ will be zero. Thus, we will just need to argue that the final misses, $\Delta_{t|t}:=T_t \setminus \hat{T}_{t|t}$, will result in bounded and small centroid observation error, $\omega_t:=p_{t,\text{obs}}-p_t$. We show how to do this in the next paragraph. This, along with ensuring the stability of $\Sigma_{t|t}$, will ensure bounded and small $|p_t - \hat{p}_{t|t}|$.
{\em Note that we use a KF in this paper, but in general the above argument will go through with any stable linear observer.}


Let $T:=\hat{T}_{t|t}$ and $\Delta:= T_t \setminus T$. To obtain a bound on $|\omega_t|$, notice that $p_t = \frac{1}{|T_t|} \sum_{i \in T_t} i$ while $p_{t,\text{obs}} = \frac{1}{|T|} \sum_{i \in T} i$.  Since the final number of extras is zero, $T_t = T \cup \Delta$ and so $|T_t| = |T|+|\Delta|$. Thus, $(|T|+|\Delta|)p_t - |T| p_{t,\text{obs}} = \sum_{i \in \Delta} i$ and so $|T|(p_t - p_{t,\text{obs}}) = \sum_{i \in \Delta} (i - p_t)$. Using (\ref{defTt}), $|T_t| = 2w+1$ and $|(i - p_t)| \le w$ for all $i \in T_t$. Thus,
\begin{equation}
|\omega_t| = |p_t - p_{t,\text{obs}}| \le \frac{|\Delta_{t|t}|w}{2w+1 - |\Delta_{t|t}|}  \label{omegabnd}
\end{equation}
In practice, if there are extras (thresholds are not set high enough to ensure zero extras), then we can show that
\begin{equation}
|\omega_t| \le |\Delta_{t|t}| \frac{w}{2w+1 - |\Delta_{t|t}|}  + |\Delta_{e,t|t}| \frac{\max_{j \in \Delta_{e,t|t}} |j-p_t|}{2w}
\label{omegabnd2}
\end{equation}

The above arguments, once formalized, will show that $S_t - \hat{S}_t$ is bounded and small. Since $\hat{L}_t  - L_t =S_t - \hat{S}_t$, the same will hold for $\hat{L}_t  - L_t$. But the above arguments require assuming that the ``noise" seen by modCS, $\beta_t = \hat{P}_{t,\perp}' L_{t}$ is bounded and small. To show this we will need to show that our recursive PCA algorithm given in the Sec \ref{sec_updatePt} is accurate enough, even when we use $\hat{L}_t$ instead of ${L}_t$. Also, we will need to assume that each element of $x_t$ (and hence of $L_t$) is bounded, e.g., follows a truncated Gaussian or uniform distribution.

\end{document}